# Foundation Artificial Intelligence Models for Health Recognition Using Face Photographs (FAHR-Face)

**Authors**: Fridolin Haugg MS[1,2]*, Grace Lee MD[1,2]*, John He BS[1], Leonard Nürnberg MS [1,2,3,4], Dennis Bontempi PhD[1,2,3,4], Danielle S. Bitterman MD[1,2], Paul Catalano ScD[5], Vasco Prudente MS[1,2,3], Dmitrii Glubokov MS[6], Andrew Warrington BS [1], Suraj Pai MS[1,2,4], Dr. Dirk De Ruysscher[7], Christian Guthier PhD[1,2], Benjamin H. Kann MD[1,2], Vadim N. Gladyshev PhD[6], Hugo JWL Aerts PhD[1,2,4,8]^, Raymond H. Mak MD[1,2]^

*Joint first authors
^Joint corresponding authors

**Affiliations:**
[1]Department of Radiation Oncology, Dana-Farber Cancer Institute/Brigham and Women's Hospital, Harvard Medical School, Boston, MA, USA.
[2]Artificial Intelligence in Medicine (AIM) Program, Mass General Brigam, Harvard Medical School, Boston, MA, USA.
[3]Radiology and Nuclear Medicine, CARIM & GROW, Maastricht University, Maastricht, The Netherlands.
[4]Department of Radiation Oncology (MAASTRO), Maastricht University, Maastricht, The Netherlands.
[5]Department of Biostatistics, Harvard T.H. Chan School of Public Health, Boston, MA, USA.
[6]Brigham and Women's Hospital, Harvard Medical School, Boston, MA, USA.
[7]Department of Radiation Oncology (MAASTRO), Maastricht University Medical Center, GROW School, Maastricht, The Netherlands.
[8]Department of Radiology, Brigham and Women's Hospital, Harvard Medical School, Boston, MA, USA

**Corresponding author:**
Raymond H. Mak, MD
Department of Radiation Oncology
Dana-Farber Cancer Institute/Brigham and Women's Hospital
Harvard Medical School
Boston, MA, 02115
Tel: (617) 617-632-5734, Fax: (617) 394-2667
rmak@partners.org

## Abstract

**Background:** Facial appearance offers a noninvasive window into health. We built FAHR-Face, a foundation model trained on >40 million facial images and fine-tuned it for two distinct tasks: biological age estimation (FAHR-FaceAge) and survival risk prediction (FAHR-FaceSurvival).

**Methods**: FAHR-FaceAge underwent a two-stage, age-balanced fine-tuning on 749,935 public images; FAHR-FaceSurvival was fine-tuned on 34,389 photos of cancer patients. Model robustness (cosmetic surgery, makeup, pose, lighting) and independence (saliency mapping) was tested extensively. Both models were clinically tested in two independent cancer patient datasets with survival analyzed by multivariable Cox models and adjusted for clinical prognostic factors.

**Findings**: For age estimation, FAHR-FaceAge had the lowest mean absolute error of 5.1 years on public datasets, outperforming benchmark models and maintaining accuracy across the full human lifespan. In cancer patients, FAHR-FaceAge outperformed a prior facial age estimation model in survival prognostication. FAHR-FaceSurvival demonstrated robust prediction of mortality, and the highest-risk quartile had more than triple the mortality of the lowest (adjusted hazard ratio 3.22; P<0.001). These findings were validated in the independent cohort and both models showed generalizability across age, sex, race and cancer subgroups. The two algorithms provided distinct, complementary prognostic information; saliency mapping revealed each model relied on distinct facial regions. The combination of FAHR-FaceAge and FAHR-FaceSurvival improved prognostic accuracy.

**Interpretation**: A single foundation model can generate inexpensive, scalable facial biomarkers that capture both biological ageing and disease-related mortality risk. The foundation model enabled effective training using relatively small clinical datasets. Routine quantitative facial analysis could enhance risk stratification when traditional biomarkers are unavailable.

## Introduction

Facial images have emerged as a promising biomarker for health assessment, offering a non-invasive and readily accessible means of capturing physiological changes associated with aging and overall health[1,2]. Influenced by various physiological factors, disease processes, and lifestyle choices, the human face may provide valuable insights into an individual's health status, reflecting the cumulative effects of genetic predisposition, environmental exposures, existing medical conditions, and lifestyle habits[3,4]. These facial changes can mirror the underlying biological aging process, connected with the development of various diseases, particularly cancer, through the time-dependent accumulation of cellular damage[5–7]. Despite this potential, effectively quantifying facial appearance for health assessment has been challenging.

Studies have demonstrated the potential of AI-based approaches to extract health-related information from facial images, including chronological age estimation[8–13], biological age estimation[14,15], cardiovascular risk assessment[14], and identification of individuals with rare genetic syndromes[2,16,17]. These approaches leverage the power of deep learning to identify subtle patterns and features in facial images that correlate with various health outcomes.

Building upon these AI advancements, foundation models have revolutionized natural language processing[18,19] and computer vision[20], and are now making substantial strides in medical applications. These models, trained on vast amounts of data, have demonstrated remarkable capabilities in capturing complex patterns and transferring knowledge across diverse tasks[21]. In medical applications, foundation models have shown promise in clinical natural language processing[22,23], medical image analysis[24], and pathology[25].

Among the techniques employed to develop such foundation models, masked autoencoders have emerged as a particularly effective approach for image-based tasks[26]. Masked autoencoders learn by randomly masking and reconstructing patches in images, creating a challenging self-supervisory task that requires a holistic understanding of the input[27–29]. This approach is especially well-suited for facial image analysis, as the face contains many correlated structures (e.g., symmetry, typical spatial relationships between features), which the model can leverage to capture higher-level attributes potentially related to health.

Here, we introduce FAHR-Face (Foundation Artificial Intelligence for Health Recognition), a deep-learning framework designed to assess health-related attributes and predict survival outcomes using consumer-grade digital photographs of the face. Our novel approach utilizes a masked autoencoder pre-trained on over 40 million facial images to develop a foundation model capable of transforming subtle facial feature variations into quantifiable imaging biomarkers, and this foundation model can then be fine-tuned for multiple health-related tasks.

Our study demonstrates the application of this foundation model in two key areas: FAHR-FaceAge for biological age estimation and FAHR-FaceSurvival for predicting survival outcomes. We demonstrate using a foundation model for pre-training enables superior biological age estimation for FAHR-FaceAge compared to state-of-the-art architectures and that the biomarkers from FAHR-FaceAge and FAHR-FaceSurvival yield distinct but complementary prognostic information in two large, independent clinical datasets of cancer patients.

**Methods**

**Study design and data**

A foundation model (FAHR-Face) was trained using self-supervised learning on over 40 million unlabeled facial images from the WebFace42M dataset[30]. Subsequently, two distinct supervised-learning models were developed by fine-tuning this foundation model: FAHR-FaceAge for biological age estimation and FAHR-FaceSurvival for predicting survival from facial photographs. The complete study design and dataset flow are illustrated in **Figure 1.**

FAHR-FaceAge was fine-tuned in a two-step training process, first using 8 oversampled, public datasets (IMDB-WIKI[9], KANFace[31], FG-NET[32], CACD[33], AFAD[34], MegaAge[35], MORPH[36], LAGENDA[37]) followed by a age-balanced fine-tuning stage on two public datasets (UTK[38] and AgeDB[39]) to ensure unbiased age estimation across human lifespan (0-116 years). The model, benchmarked against state-of-the-art architectures[40] (APPA-REAL[41]) and assessed for robustness to image variations[42–44], was evaluated on the entire Harvard RT cohort (N=38,211) and externally validated on the independent Maastro cohort (N=4,892).

FAHR-FaceSurvival was developed by fine-tuning FAHR-Face specifically on a subset of the Harvard RT cohort. The Harvard RT images were randomly partitioned (90:10 split), resulting in a training set of 34,389 images and a held-out test set of 3,822 images. Training employed a ranking-based loss approach to predict survival outcomes. The final FAHR-FaceSurvival model performance was evaluated on the held-out test set and independently validated using the same Maastro cohort. See **Supplementary Materials** for additional details on model training.

**End-points**

Chronological-age accuracy was summarized with mean absolute error (MAE; overall and averaged across 5-year bins). Biological ageing was expressed as FAD (FAHR-FaceAge deviation =FAHR-FaceAge minus chronological age). Survival discrimination was assessed with Harrell's C-index, multivariable Cox models, Kaplan-Meier curves and time-dependent AUCs (3, 6, 12, 24 months). All primary/secondary endpoints were validated in the external Maastro cohort (N=4892).

## Statistical Analyses

Baseline characteristics were summarized descriptively. Survival was modelled with Cox proportional hazards; multivariable models included age (per decade), sex, race/ethnicity, cancer site, treatment intent, calendar year and radiation technique, retaining covariates with P<0.05 in univariate screening. Discrimination was quantified with Harrell's C-index and time-dependent AUCs at 3, 6, 12 and 24 months, comparing chronological age, FAD and FAHR-FaceSurvival. Risk stratification used Kaplan-Meier curves and log-rank tests; subgroup hazard ratios with 95% CIs were reported. Complementarity and independence of the two models were assessed in both feature and geometric space. Analyses used R 4.3.1 and SAS 9.4; two-sided P<0.05 was considered significant.

## Ethical approval

Institutional review board approval with waiver of consent was obtained for this retrospective analysis, and all identifiable facial images were handled in accordance with institutional and ethical guidelines.

## Role of the funding source

The funders of the study had no role in the study design, data collection, data analysis, data interpretation, or writing of the report.

## Results

### A. FAHR-FaceAge (Biological Age Estimation Model)

The FAHR-FaceAge model for biological age estimation, which was fine-tuned from the FAHR-Face foundation model, achieved the lowest mean absolute error (MAE) of 5.1 years on the public APPA-Real dataset of presumed healthy individuals **(Figure 2a; Extended Data Figure 1a,c).** In comparison as a benchmark**,** a baseline ViT trained from random initialization reached 6.92 years MAE (see Supplementary materials for additional benchmark details and results). Fine-tuned with age-balanced data to reduce bias, FAHR-FaceAge demonstrated generalizability across diverse age ranges and racial groups in publicly available, real-world photographs and photographs obtained in clinical settings **(Figure 2b)**. The heatmaps reveal consistent age estimation accuracy across racial subgroups with high correlations (White: 0.86, Black: 0.81, Asian: 0.86, Hispanic: 0.91, Other: 0.91).

Next, we tested FAHR-FaceAge as a biological age biomarker in clinical cohorts and demonstrated its superiority to a previously published age estimation algorithm (FaceAge[15]). The clinical application of the FAHR-FaceAge model was first evaluated by analyzing the FAHR-FaceAge deviation (FAD). FAD is the difference between the estimated biological age from the FAHR-FaceAge model and the patient's chronological age (FAD = FAHR-FaceAge minus Chronological Age). FAD was tested as a clinical biomarker on a large independent cohort of 38,211 patients with a diverse range of cancer diagnoses (Harvard RT Patient Dataset). The median FAD was 1.1 years, with 33.0% of patients having a FAD ≥ 5 years and 17.1% having a FAD ≥ 10 years, i.e., patients appearing much older than their chronological age. Conversely, 24.3% had a FAD ≤ -5 years, i.e., patients appearing much younger than their chronological age. The median follow-up time in this cohort was 2.1 years[45]. The patient characteristics are summarized in **Extended Data Table 1**.

Patients with higher FADs exhibited lower survival probabilities across all chronological age groups, with significant differences observed between patients with FAD differences exceeding 5 years or more (**Figure 2e**); patients appearing younger than their chronological age showed progressively higher survival rates, while those appearing older had increasingly poorer outcomes. This association between FAD and survival is evident across all age and race subgroups (**Figures 2g and 2h**). Additionally, FAD is a robust early mortality biomarker (**Figure 2f**); Patients in the highest FAD group (>20 years) had a 90-day mortality of 21.4%, compared to less than 5% for those with FAD ≤ -10 years. These findings underscore FAD's potential as an effective biomarker of long-term survival and early mortality in cancer patients.

Benchmarking against the previously published biological face age estimation model (FaceAge[15]), in single-predictor analyses, our new FAD biomarker achieved a C-index of 0.57 while the prior model's age difference biomarker (FaceAge minus Age) achieved 0.52, with FAD maintaining higher C-indices across age and racial subgroups (**Figure 2c**). Examining different survival endpoints in patients aged 60 and older, FAHR-FaceAge had higher performance for 3 months, 6 months, 1 year, and 2 year survival prediction than FaceAge and chronological age **(Figure 2d).** These findings demonstrate that using a foundation model to inform a biological face age estimation model can improve prognostic performance compared to our previously published model, which was trained in a supervised, task-specific manner.

In univariate Cox regression analyses **(Table 1)**, FAD (per decade) emerged as a significant predictor of survival (HR = 1.30, 95% CI 1.27–1.32, P<0.001). Multivariate analysis adjusting for age, sex, race/ethnicity, cancer site, treatment intent, treatment year, and radiation technique confirmed FAD's significance (adjusted HR = 1.14, 95% CI 1.12–1.16, P<0.001). Notably, a FAD ≥ 5 years (vs < 5) was linked to a 21% higher hazard (adjusted HR 1.21, 95 % CI 1.17–1.26, P<0.001). Conversely, a FAD ≤ −5 years predicted a 17% lower hazard compared with those whose FAD > −5 years (vs ≤ −5; adjusted HR 0.83, 95 % CI 0.79–0.87, P<0.001).

Subgroup analyses **(Extended Data Table 2)** were generally consistent. While both males and females exhibited significant FAD-survival associations, females showed slightly higher hazard ratios. White and Asian patients had stronger FAD-based risk stratification than Black and Hispanic groups. Across cancer types, patients with FAD ≥ 5 years (vs < 5) consistently had poorer survival **(Extended Data Figure 2)**.

Lifestyle factors were significantly associated with FAD: both heavy alcohol consumption and drug use corresponded with increased FAD (P<0.05). Notably, while FAHR-FaceAge detected significant associations with alcohol use in both men and women, the previous FaceAge[15] model failed to detect significant associations with alcohol use in women. The strongest association was observed with smoking, showing consistent dose-dependent relationships across both smoking-related and non-smoking-related cancers: both in smoking status (current > former > never smokers) and lifetime exposure (moderate-heavy > light > never smokers) (P<0.001; **Extended Data Figure 3)**.

These findings demonstrate that FAHR-FaceAge surpasses existing performance in both chronological and biological age estimation across populations while capturing clinically meaningful health variations that reflect both disease severity and lifestyle-related physiological changes.

## B. FAHR-FaceSurvival (Risk Estimation Model)

Next, we independently trained the FAHR-FaceSurvival model from patient photographs by fine-tuning FAHR-Face foundation model for survival prediction. The model was internally validated on a sub-cohort of 3,822 patients from the Harvard RT Test Dataset. In the test cohort, the median age was 64.1 years (range 0.8 to 98.5). The test cohort was 57.0% female, with a racial distribution of 87.6% White, 4.9% Black, 3.1% Asian, and 4.4% other races. Most patients (68.9%) had non-metastatic cancer, 27.3% had metastatic cancer, and 3.8% had benign conditions. The median follow-up time was 2.0 years[45]. The training and test patient characteristics are summarized in **Extended Data Table 1.**

Higher FAHR-FaceSurvival scores were associated with more advanced/aggressive cancer risk groups, with a median of 0.125 for benign/DCIS conditions, 0.244 for primary (non-metastatic) cancer, and 0.427 for metastatic cancer (P<0.001; **Figure 3a**). Among age groups, FAHR-FaceSurvival predicts the highest mortality risk in the 60-80 and 80+ age groups (**Figure 3a**). The FAHR-FaceSurvival was also significantly higher in females (0.357 for females vs 0.214 for males; P<0.001), with no significant difference between most races (P > 0.05 for all comparisons except Asian vs White, P=.028; **Figure 3a**). Higher FAHR-FaceSurvival quartiles were associated with significantly lower survival probabilities (**Figure 3b**).

On univariate Cox regression, the continuous FAHR-FaceSurvival was associated with survival (HR for every 0.1 increase = 1.32, 95% CI 1.30-1.35, P<0.001). Multivariate analysis, adjusting for age, sex, race/ethnicity, cancer site, treatment course intent, year treated, and radiation technique, confirmed that FAHR-FaceSurvival remained significantly associated with survival (adjusted HR = 1.17, 95% CI 1.15-1.20, P<0.001; **Table 2**).

Subgroup analysis demonstrated FAHR-FaceSurvival association with survival across various demographic and clinical subgroups. As shown in **Extended Data Table 3**, both continuous (per 0.1 increase) and dichotomized (≥ 0.5 vs. < 0.5) FAHR-FaceSurvival was significantly associated with survival in univariate and multivariate analyses for most subgroups. Kaplan-Meier plots of dichotomized FAHR-FaceSurvival for different racial and age subgroups are shown in **Figure 3e** and **Figure 3f**, respectively. FAHR-FaceSurvival as a continuous variable (per 0.1 increase) showed significant associations with survival in all age and most racial subgroups, including for White (adjusted HR: 1.18, 95% CI: 1.15-1.21, P<0.001), Black (adjusted HR: 1.27, 95% CI: 1.14-1.42, P<0.001), and Asian (adjusted HR: 1.29,95% CI: 1.14-1.47, P<0.001) patients (**Extended Data Table 3**). However, the association was not statistically significant for Hispanic patients (HR: 5.35, 95% CI: 0.93-30.73, P=.060).

FAHR-FaceSurvival was also significantly associated with early mortality risk. Over 50% of patients in the highest FAHR-FaceSurvival (0.9-1.0) group died within 90 days, compared to only 2% among the lowest FAHR-FaceSurvival (0.0-0.1) group **(Figure 3d)**. The C-index for race, age, and diagnosis subgroups, showing varied performance with an overall C-index of 0.72 and ranging from 0.53 to 0.88 across different subgroups (**Figure 3g**). Notably, a benchmark model trained on cancer patient photographs alone without the FAHR-Face foundation model using a ViT with random-initialization had a C-index of 0.66 (**see Supplementary materials**), illustrating the value of fine-tuning FAHR-FaceSurvival from the foundation model.

These findings demonstrate that FAHR-FaceSurvival can robustly identify patients at high risk of early mortality across diverse patient subgroups.

## C. External Validation of FAHR-Face-Derived Models in an Independent Patient Cohort

The generalizability of both FAHR-Face models was validated using 4,892 patients from the Maastro Radiotherapy (RT) clinic. This cohort had a median age of 66.0 years, with 47.2% female patients. Most common cancer sites included breast, gastrointestinal, genitourinary, and lung (**Extended Data Table 1**). The median follow-up time was 3.0 years.

The median FAHR-FaceAge was 70.9 years (range: 12.0-97.0), with a median FAD of 4.4 years (range: -31.0 to 39.6). 46.8% of patients had a FAD ≥ 5 years, and 22.5% had a FAD ≥ 10 years. The MAE for FAHR-FaceAge was 6.99 years, and the ME was 4.70 years.

The performance of FAHR-FaceAge as a prognostic factor for survival was validated in this cohort. Kaplan-Meier survival analysis revealed significant stratification based on FAD **(Figure 4a)**. Multivariable analysis, adjusted for age, sex, and cancer site, confirmed FAD as a significant predictor of survival (adjusted HR = 1.127, 95% CI 1.057-1.201, P<0.001; **Figure 4c)**.

For FAHR-FaceSurvival, the median was 0.4 in this cohort. Kaplan-Meier analysis showed significant survival differences among FAHR-FaceSurvival groups **(Figure 4b)**. Multivariable analysis, adjusted for age, sex, and cancer site, confirmed FAHR-FaceSurvival (per 0.1 unit) as a significant predictor of survival (adjusted HR = 1.086, 95% CI 1.062-1.110, P<0.001; **Figure 4d)**.

The analysis of the Area under the Curve (AUC) for patients aged 60 and above **(Figure 4g)** demonstrated that FaceAge[15] and FAHR-FaceAge outperformed chronological age in predicting survival outcomes. At 3 months, FAHR-FaceAge achieved an AUC of 0.66 compared to FaceAge of 0.64 and chronological age of 0.63. This performance advantage persisted at 6 months, 1 year, and 2 years.

This independent external validation in a geographically distinct cohort demonstrates the generalizability of both FAHR-Face models across different clinical settings and patient populations.

## D. FAHR-FaceAge and FAHR-FaceSurvival Robustness, Interpretability, and Model Independence

FAHR-FaceAge model robustness was evaluated using paired images of the same person captured within short timespans from three distinct public datasets, including 1) before and after plastic surgery; 2) with and without makeup; and 3) varying facial expressions and photograph illumination and angles **(Extended Data Figure 1e-g)**. Age estimates were not significantly different for eyebrow correction, eyelid correction, facelifts, or facial bone correction, but were higher after nose correction (mean increase 1.11 years, P<0.001). No statistical difference was seen with makeup, viewing angle, illumination, or facial expression.

Attention maps were generated from the trained models to understand which regions of the face contribute most to each individual model's predictions in all patients from the Harvard RT Test (n=3,822). The attention maps for FAHR-FaceSurvival were localized to regions below the eyes and including the nasal bridge, as visualized in **Figure 3c,** and 3D projections onto the canonical face model in **Extended Data Figure 4b,** indicating that these areas are important for predicting patient survival outcomes. In contrast, the FAHR-FaceAge attention maps (**Extended Data Figure 4a**) predominantly highlighted the nasolabial folds, forehead, and temporalis area, underlining the importance of age-related changes in these areas. Age-stratified attention maps **(Extended Data Figure 4c)** illustrate further the distinct salient features for each model: for FAHR-FaceAge, the regions with high levels of attention shift over time, including a heavier emphasis on the temporalis and forehead areas in older age groups (40 years and onwards), while FAHR-FaceSurvival maintains relatively consistent attention patterns across all age groups.

The independence of the two models is further quantitatively supported by a low correlation between FAD and FAHR-FaceSurvival (R = 0.21**; Extended Data Figure 4d**) and minimal feature space overlap (median cosine similarity: -0.018, **Extended Data Figure 4e**) in the Harvard RT Test Dataset. FAHR-FaceAge and FAHR-FaceSurvival showed a low correlation (R = 0.160) **(Figure 4e)**, demonstrating that the models capture different aspects of patient health from face photographs in the independent Maastro RT cohort despite a shared foundation model.

In a combined Cox model, both FAD (HR=1.074, 95% CI 1.007-1.146, P=0.03) and FAHR-FaceSurvival (HR=1.080, 95% CI 1.055-1.104, P<0.001) remained independent predictors of survival **(Figure 4f)**. Separate Cox models were fitted for FAD and FAHR-FaceSurvival individually, as well as for a combined model, each adjusted for age, sex, and cancer site. The combined model exhibited a lower (AIC=26560) relative to the FAD (AIC=26602) and FAHR-FaceSurvival models (AIC=26563), indicating superior model fit. These results reinforce that the biomarkers provide complementary prognostic information for overall survival.

These analyses demonstrate that FAHR-Face models are robust to common image variations, derive predictions from anatomically distinct facial regions, and capture complementary aspects of health status.

**Discussion**

Our study advances non-invasive health assessment significantly by quantitatively analyzing facial images using a foundation model trained on over 40 million unlabeled facial photographs. Leveraging masked autoencoders and self-supervised learning, our FAHR-FaceAge model demonstrated superior performance and generalizability in biological age estimation compared to previous models[13], and we demonstrated that the foundation model can enable training of a clinical outcomes-specific facial health algorithm, FAHR-FaceSurvival, overcoming limitations related to clinical dataset size and generalizability. Unlike prior models relying on specialized imaging[14], our approach effectively utilizes low-cost, easily acquired consumer-grade facial photographs, significantly broadening clinical applicability.

Our findings suggest that facial features encode both physiological aging and imminent mortality risk, demonstrating that consumer-grade photographs carry multiple physiological health signals. Despite their shared backbone, FAHR-FaceAge and FAHR-FaceSurvival offer non-overlapping prognostic information, as supported by saliency mapping and multivariable analyses. Saliency maps demonstrated their independence; FAHR-FaceSurvival concentrates on the area under the eyes and around the nose, regions of the face linked with dying[46], while FAHR-FaceAge maps to nasolabial folds and temples, which are regions associated with aging[47,48]. Furthermore, the clinical use cases are complementary as well: FAD, given in years, is intuitive and can track aging or be incorporated into clinical risk stratification tools, while FAHR-FaceSurvival may be best suited for near-term mortality prediction.

The growing elderly population makes rapid frailty assessment essential for treatment planning[49,50]. FAHR-Face models could complement or enhance traditional risk calculators, especially where formal assessments (e.g., frailty scales, grip strength or the Timed Up and Go test)[51–53] are impractical. A single facial photograph analyzed by our FAHR-Face models provides a non-invasive, inexpensive alternative that can flag patients unlikely to tolerate aggressive therapy, improving resource allocation and quality of life[54,55]. As facial features also reflect behaviors such as smoking and alcohol use[56–58], the same tool can track lifestyle change or anti-aging interventions, offering immediate feedback even in resource-limited settings.

Limitations include cancer-only cohorts from mainly Western centers, which may limit transferability to other diseases, regions, and under-represented racial groups. Image quality, lighting, and facial expression can still affect accuracy, highlighting the value of standardized photo capture[13]. Finally, regulatory, workflow, and reimbursement hurdles must be cleared before facial biomarkers can enter routine care[59,60].

Using identifiable facial images demands strict privacy safeguards. Bias remains a risk, so we audited performance across race and age, yet wider representation is still required[61]. Clear regulation and governance are vital to prevent misuse and ensure equitable deployment[62–65]. Next steps include validation in broader cohorts, integration with other biomarkers, and prospective trials to confirm clinical benefit while developing practical capture standards and clinician training[67–69].

In conclusion, our study demonstrates that quantifying imaging biomarkers from facial photographs through a foundation model approach offers a transformative and accessible method for assessing overall health status and predicting survival outcomes. The foundation model approach provides the ability to train multiple independent models to predict health outcomes, even when using relatively limited in-size clinical datasets, which will broaden the potential clinical applications of facial health assessments.

**Data availability**

The WebFace42M dataset used for foundation model training is available at https://www.face-benchmark.org/download.html. The public datasets used for FAHR-FaceAge model development and evaluation are available through their respective repositories: IMDB-WIKI (https://data.vision.ee.ethz.ch/cvl/rrothe/imdb-wiki/), KANFace (https://ibug.doc.ic.ac.uk/resources/kanface/), FG-NET (https://yanweifu.github.io/FG_NET_data/), CACD (https://bcsiriuschen.github.io/CARC/), (https://github.com/John-niu-07/tarball/), MegaAge (http://mmlab.ie.cuhk.edu.hk/projects/MegaAge/), MORPH (https://uncw.edu/myuncw/research/innovation-commercialization/technology-portfolio/morph/), LAGENDA (https://wildchlamydia.github.io/lagenda/), UTK (https://susanqq.github.io/UTKFace/), AgeDB (https://ibug.doc.ic.ac.uk/resources/agedb/), and APPA-REAL (http://chalearnlap.cvc.uab.es/dataset/26/description/). The datasets used for robustness evaluation are available at: HDA Plastic Surgery Database (https://dasec.h-da.de/research/biometrics/hda-plastic-surgery-face-database/), Makeup Dataset (https://www.kaggle.com/datasets/tapakah68/makeup-detection-dataset), and FEI Face Database (https://fei.edu.br/~cet/facedatabase.html).

Clinical data from the Harvard RT Patient Dataset and Maastro RT Dataset contain protected health information and cannot be made publicly available due to patient privacy regulations.

## Contributors

FH and GL contributed to conceptualization, data curation, formal analysis, investigation, methodology, software, validation, visualization, writing – original draft, and writing – review & editing. JH, LN, DB, DSB, PC, VP, DG, AW, SP, DDR, CG, BK, and VG contributed to data curation, formal analysis, investigation, methodology, validation, visualization, and writing – review & editing. HJWLA and RHM contributed to conceptualization, funding acquisition, project administration, resources, supervision, writing – review & editing, and methodology.

## Code availability

The code used to train the FAHR-Face model, FAHR-FaceAge, and FAHR-FaceSurvival models, will be made available on GitHub upon publication. The repository will include preprocessing pipelines, model architectures, training procedures, with documentation. Additionally, the trained weights for FAHR-Face and FAHR-FaceAge model will also be publicly available.

## Conflicts of Interest:

RHM reports being on an Advisory Board for ViewRay and AstraZeneca; Consulting for AstraZeneca, Varian Medical Systems, and Sio Capital Management; Honorarium from Novartis and Springer Nature; and Research Funding from the National Institute of Health, ViewRay, AstraZeneca, Siemens Medical Solutions USA, Inc., and Varian Medical Systems. DSB reports Editorial, unrelated to this work: Associate Editor of Radiation Oncology, HemOnc.org (no financial compensation); Research funding, unrelated to this work: American Cancer Society, American Society for Radiation Oncology, National Institutes of Health; Advisory and consulting, unrelated to this work: MercurialA. HJWLA reports advising and consulting for Onc.AI, Love Health, Sphera, Editas, AstraZeneca, and Bristol Myers Squibb, unrelated to this work. BHK reports research funding from the Botha-Chan Low Grade Glioma Consortium (National Institutes of Health [NIH]-USA K08DE030216-01). DR report institutional financial interests (no personal financial interests) for AstraZeneca (unrestricted research grants and advisory board), Bristol-Myers-Squibb (unrestricted research grants and advisory board), Siemens (unrestricted research grants), Beigene (unrestricted research grant), Philips (research support and advisory board), Olink (research support) and Eli-Lilly (support). All other authors declare no competing interests. FH, RHM, and HJWLA report that Mass General Brigham has filed provisional patents on two next-generation facial health algorithms. All other authors declare no competing interests.

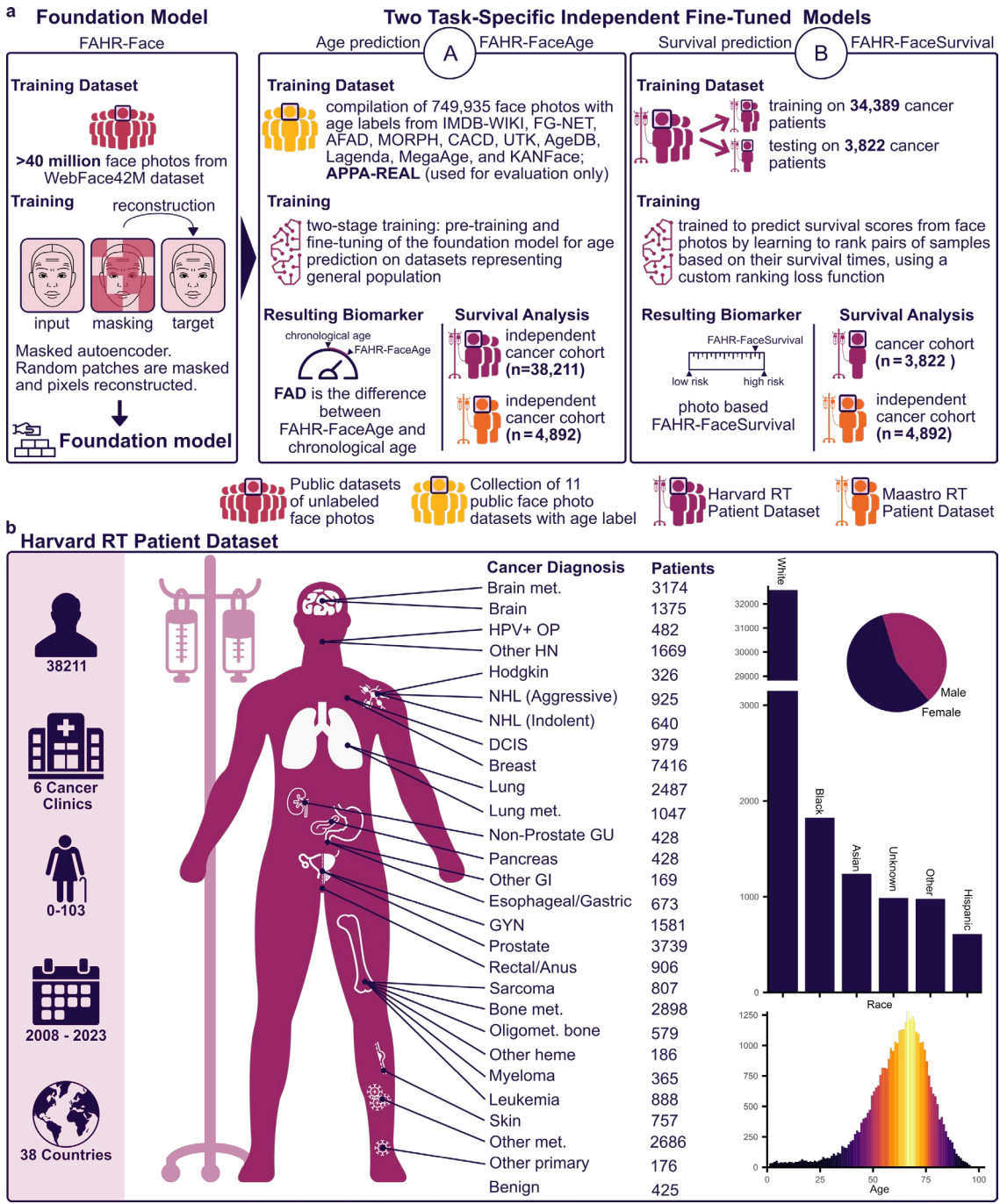

**Figure 1: Study Design and Harvard RT Patient Dataset Overview**

**a,** Study design flowchart. The FAHR-Face foundation model is trained on more than 40 million face photos from the WebFace42M dataset. Two applications are derived: FAHR-FaceAge for age prediction and FAHR-FaceSurvival for survival prediction. FAHR-FaceAge uses 749,935 face photos with age labels from 11 public datasets and is evaluated on the independent Harvard RT dataset (n = 38,211). FAHR-FaceSurvival is trained on 34,389 patients from the Harvard RT dataset and tested on 3,822 patients (90/10 split). Both models are evaluated on the independent Maastro RT dataset (n = 4,892). FAD, FAHR-FaceAge deviation (FAHR-FaceAge minus chronological age); RT, radiotherapy.

**b,** Overview of the Harvard RT Patient Dataset. 38,211 patients treated with RT from 2008 to 2023. The dataset includes various cancer types and patient demographics. The pie chart shows sex distribution. The bar chart displays race distribution. The histogram illustrates the age distribution. RT, radiotherapy; met., metastases; HPV, human papillomavirus; OP, oropharyngeal; HN, head and neck; NHL, non-Hodgkin lymphoma; DCIS, ductal carcinoma in situ; GU, genitourinary; GI, gastrointestinal; GYN, gynecological.

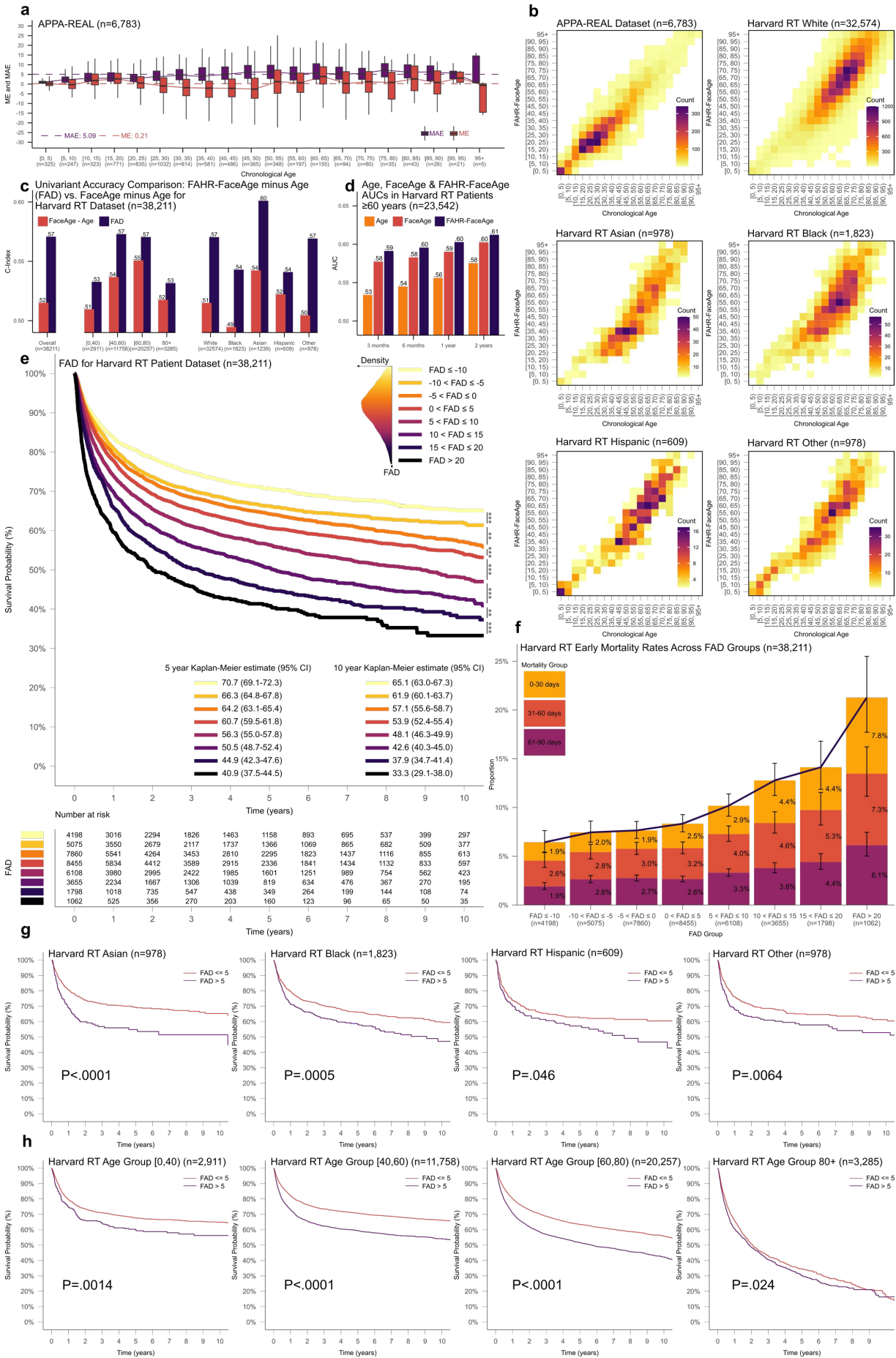

**Figure 2: FAHR-FaceAge and FAD Analysis in a Clinical Dataset (Harvard RT) and a Publicly Available Dataset (APPA-Real)**

**a,** Mean Absolute Error (MAE) and Mean Error (ME) per 5-year age bin for age estimation on the public APPA-REAL dataset. This panel demonstrates consistent performance across all age bins, indicating that the model's accuracy in predicting age is stable throughout different age groups.

**b,** Predicted age versus actual age heatmaps for different race groups in the Harvard RT Patient and APPA-REAL datasets (correlation: 0.92). The heatmaps illustrate that while different race groups exhibit different age distributions, the model shows similar diagonal patterns, with high correlations across all groups (White: 0.86, Black: 0.81, Asian: 0.86, Other: 0.91, Hispanic: 0.91).

**c,** Benchmarked in single-predictor Cox models, the difference between FAHR-FaceAge and chronological age consistently outperformed FaceAge across overall population and stratifications by age and race groups, as measured by unadjusted C-indices. Sample sizes are shown for each subgroup.

**d,** Area Under the Curve (AUC) comparison of Age, FaceAge, and FAHR-FaceAge for survival prediction in Harvard RT patients aged 60 and above (n = 23,542). AUC values were calculated at four time points: 3 months, 6 months, 1 year, and 2 years. FAHR-FaceAge consistently demonstrated superior predictive performance across all time points, followed by FaceAge, with both outperforming chronological age. The analysis was restricted to patients over 60 years old because the FaceAge[15] model was only trained and validated for individuals above 60 years of age.

**e,** Kaplan-Meier survival curves stratified by FAD (FAHR-FaceAge minus chronological age) groups for the Harvard RT Patient dataset. The FAD groups range from -10 to 20+ years. The curves show that higher FAD values are associated with poorer survival outcomes, demonstrating that looking older than one's chronological age correlates with shorter survival times. (log-rank P-values: ***P ≤ 0.001, **P ≤ 0.01, *P ≤ 0.05). The sample density plot shows the relative distribution of patients per FAD group.

**f,** Percentage of patients with early mortality (within 30, 31-60, and 61-90 days) across different FAD groups. The stacked plot represents a continuous increase in the proportion of patients who die within 90 days as the FAD value increases. Error bars indicate 95% confidence intervals.

**g,** Kaplan-Meier survival curves for the Harvard RT Patient dataset, split by FAD > 5 and FAD ≤ 5, across different race groups (Asian, Black, Hispanic, and Other). The significant differences in survival between the groups highlight the robustness of FAD as a biomarker across different racial demographics (log-rank P-values).

**h,** Kaplan-Meier survival curves for the Harvard RT Patient dataset, split by FAD > 5 and FAD ≤ 5, across different age groups (0-40, 40-60, 60-80, and 80+). The significant differences in survival between the groups demonstrate the utility of FAD in predicting survival outcomes across various age demographics (log-rank P-values).

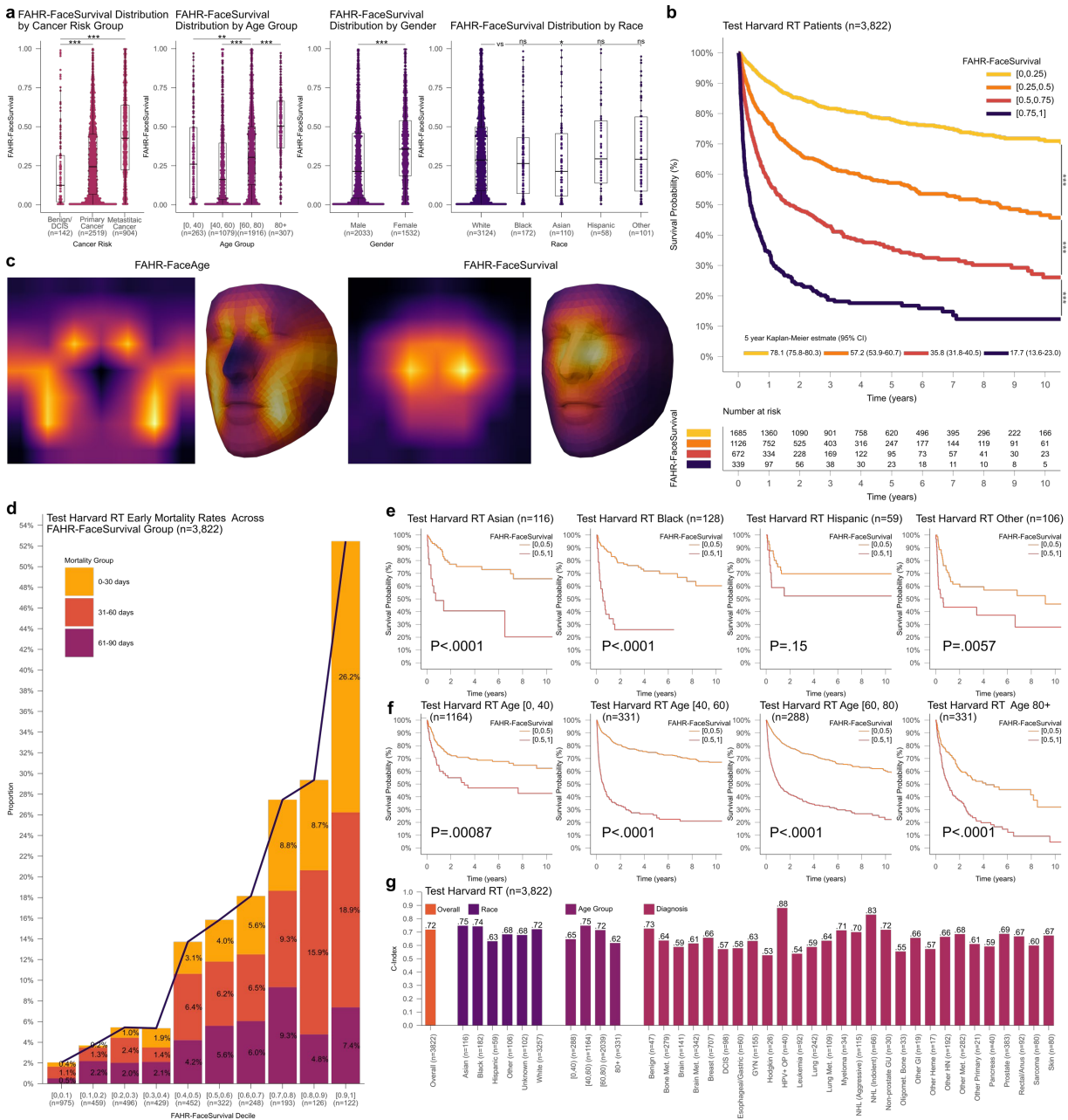

**Figure 3**: **FAHR-FaceSurvival Analysis on the Harvard RT Test Dataset**

**a**, Box plots show the variability of FAHR-FaceSurvival across different demographic groups, indicating that the model captures meaningful variations in risk based on these factors (Wilcoxon rank-sum test: ns (P>0.05), * (P≤0.05), ** (P≤0.01), *** (P≤0.001).

**b**, Risk stratification by grouping patients into quartiles based on their FAHR-FaceSurvival. The Kaplan-Meier survival curves show significant differences in survival probabilities between the quartiles, indicating the model's ability to effectively stratify patients by risk (log-rank P-values: ***P≤0.001, **P≤0.01, *P≤0.05).

**c,** Mean attention maps for FAHR-FaceAge and FAHR-FaceSurvival, showing the facial regions most relevant for each prediction. Results are displayed on both 2D projections and 3D canonical face models.

**d**, Early mortality percentages within 30, 31-60, and 61-90 days across different FAHR-FaceSurvival groups, divided into deciles (1/10 steps). The stacked plot shows a continuous increase in early mortality rates with higher FAHR-FaceSurvival.

**e**, Kaplan-Meier survival curves for the Harvard RT Patient dataset, split by FAHR-FaceSurvival above and below 0.5, across different race groups (White, Black, Asian, Hispanic, and Other). Significant differences in survival between the groups underscore the robustness of the FAHR-FaceSurvival across various racial demographics (log-rank P-values).

**f**, Kaplan-Meier survival curves for the Harvard RT Patient dataset, split by FAHR-FaceSurvival above and below 0.5, across different age groups (0-40, 40-60, 60-80, and 80+). Significant differences in survival between the groups demonstrate the utility of FAHR-FaceSurvival in predicting survival outcomes across various age demographics (log-rank P-values).

**g**, Concordance Index (C-index) for overall age group, sex, race, and cancer types. This panel demonstrates the performance of the FAHR-FaceSurvival model as a single predictor in estimating overall survival across various subgroups, with higher C-index values indicating better predictive accuracy.

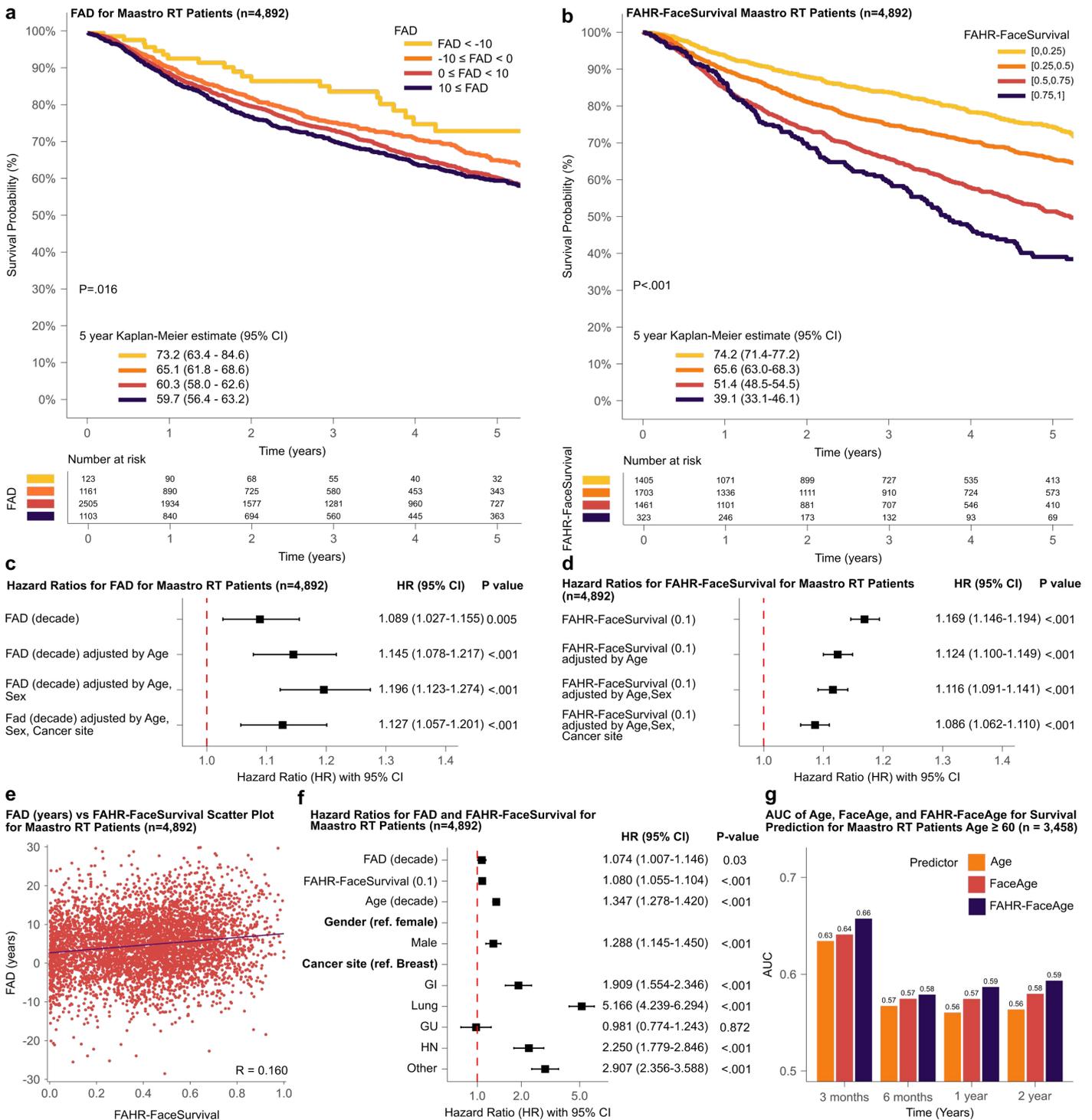

**Figure 4: Prognostic performance of FAD and FAHR-FaceSurvival on the External Validation Dataset.**

**a,** Kaplan-Meier survival curves stratified by FAD groups. Patients with lower FAD values demonstrated higher survival probabilities. The 2.5-year and 5-year Kaplan-Meier estimates (95% CI) are provided for each group. FAD, FAHR-FaceAge difference (FAHR-FaceAge minus chronological age); CI, confidence interval.

**b,** Kaplan-Meier survival curves stratified by FAHR-FaceSurvival. Lower FAHR-FaceSurvival was associated with higher survival probabilities. The 2.5-year and 5-year Kaplan-Meier estimates (95% CI) are shown for each group. CI, confidence interval.

**c,** Forest plot showing hazard ratios (HR) with 95% confidence intervals for FAD (per decade) in univariate and multivariate analyses. FAD remained a significant predictor of survival after adjusting for age, sex, and cancer site. FAD, FAHR-FaceAge deviation (FAHR-FaceAge minus chronological age); CI, confidence interval.

**d,** Forest plot displaying hazard ratios (HR) with 95% confidence intervals for FAHR-FaceSurvival in univariate and multivariate analyses. FAHR-FaceSurvival maintained significant prognostic value after adjusting for age, sex, and cancer site.

**e,** Scatter plot illustrating the relationship between FAD (years) and FAHR-FaceSurvival. The low correlation (R = 0.160) suggests that these measures capture different aspects of patient health. FAD, FAHR-FaceAge deviation (FAHR-FaceAge minus chronological age); CI, confidence interval.

**f,** Forest plot comparing hazard ratios (HR) with 95% confidence intervals for FAD, FAHR-FaceSurvival, age, sex, and cancer sites in a multivariate analysis. Both FAD and FAHR-FaceSurvival remained significant predictors when considered in the same Cox model. FAD, FAHR-FaceAge deviation (FAHR-FaceAge minus chronological age); CI, confidence interval; ref., reference.

**g,** Area Under the Curve (AUC) comparison of Age, FaceAge, and FAHR-FaceAge for survival prediction in MAASTRO RT patients aged 60 and above (n = 3,458). AUC values were calculated at four time points: 3 months, 6 months, 1 year, and 2 years. FAHR-FaceAge consistently demonstrated superior predictive performance across all time points, followed by FaceAge, with both outperforming chronological age. The analysis was restricted to patients over 60 years old because the FaceAge[15] model was only trained and validated for individuals above 60 years of age.

**Table 1. Univariate and multivariate Cox regression for overall survival – FAHR-FaceAge**

| Cox regression | Univariate[a] | | Multivariate model 1[b] ([1]FAD [decade]) | | Multivariate model 2[b] ([2]FAD ≥5 years) | | Multivariate model 3[b] ([3]FAD ≤-5 years) | |
|---|---|---|---|---|---|---|---|---|
| Variables | HR (95% CI) | P value | HR (95% CI) | P value | HR (95% CI) | P value | HR (95% CI) | P value |
| Age (decade) | 1.19 (1.18-1.21) | <0.001 | 1.14 (1.13-1.16) | <0.001 | 1.14 (1.13-1.16) | <0.001 | 1.15 (1.14-1.17) | <0.001 |
| FAHR-FaceAge (decade) | 1.20 (1.18-1.21) | <0.001 | -- | -- | -- | -- | -- | -- |
| [1]FAD (decade) | 1.30 (1.27-1.32) | <0.001 | 1.14 (1.12-1.16) | <0.001 | -- | -- | -- | -- |
| [2]FAD ≥5 vs <5 | 1.53 (1.48-1.58) | <0.001 | -- | -- | 1.21 (1.17-1.26) | <0.001 | -- | -- |
| [3]FAD ≤-5 vs >-5 | 0.68 (0.66-0.71) | <0.001 | -- | -- | -- | -- | 0.83 (0.79-0.87) | <0.001 |
| Sex, male vs female | 1.38 (1.34-1.43) | <0.001 | 1.12 (1.08-1.16) | <0.001 | 1.13 (1.09-1.17) | <0.001 | 1.10 (1.06-1.14) | <0.001 |
| Race/ethnicity[c] | | | | | | | | |
| White | Ref | -- | Ref | -- | Ref | -- | Ref | -- |
| Black | 0.90 (0.83-0.98) | 0.014 | 1.11 (1.02-1.20) | 0.021 | 1.08 (0.99-1.18) | 0.073 | 1.11 (1.02-1.21) | 0.019 |
| Asian | 0.83 (0.75-0.92) | <0.001 | 1.04 (0.93-1.16) | 0.480 | 1.00 (0.89-1.11) | 0.933 | 1.02 (0.92-1.14) | 0.723 |
| Hispanic | 1.01 (0.88-1.15) | 0.937 | 1.22 (1.07-1.40) | 0.004 | 1.20 (1.05-1.37) | 0.008 | 1.21 (1.05-1.38) | 0.007 |
| Other | 0.96 (0.86-1.07) | 0.470 | 1.14 (1.02-1.28) | 0.022 | 1.14 (1.02-1.28) | 0.024 | 1.15 (1.03-1.29) | 0.013 |
| Cancer risk group[d] | | | | | | | | |
| Non-metastatic cancer | Ref | -- | -- | -- | -- | -- | -- | -- |
| Benign/DCIS | 0.18 (0.15-0.23) | <0.001 | -- | -- | -- | -- | -- | -- |
| Metastatic cancer | 5.76 (5.56-5.96) | <0.001 | -- | -- | -- | -- | -- | -- |
| Cancer site | | | | | | | | |
| Breast | Ref | -- | Ref | -- | Ref | -- | Ref | -- |
| Benign | 2.42 (1.82-3.21) | <0.001 | 2.71 (2.02-3.63) | <0.001 | 2.74 (2.04-3.67) | <0.001 | 2.72 (2.03-3.64) | <0.001 |
| Prostate | 1.67 (1.44-1.93) | <0.001 | 1.73 (1.46-2.05) | <0.001 | 1.75 (1.47-2.07) | <0.001 | 1.75 (1.47-2.07) | <0.001 |
| Lung | 15.40 (13.82-17.16) | <0.001 | 13.71 (12.17-15.45) | <0.001 | 13.95 (12.38-15.71) | <0.001 | 14.08 (12.50-15.86) | <0.001 |
| Gastrointestinal | 10.72 (9.56-12.01) | <0.001 | 9.86 (8.71-11.15) | <0.001 | 9.91 (8.76-11.22) | <0.001 | 10.00 (8.84-11.31) | <0.001 |
| Gynecological | 5.47 (4.75-6.29) | <0.001 | 6.69 (5.74-7.80) | <0.001 | 6.73 (5.78-7.85) | <0.001 | 6.74 (5.78-7.86) | <0.001 |
| Head and neck | 5.54 (4.88-6.28) | <0.001 | 5.87 (5.09-6.77) | <0.001 | 5.92 (5.13-6.83) | <0.001 | 5.98 (5.18-6.90) | <0.001 |
| Sarcoma | 6.26 (5.33-7.35) | <0.001 | 6.90 (5.81-8.18) | <0.001 | 6.93 (5.84-8.22) | <0.001 | 6.91 (5.82-8.20) | <0.001 |
| Hematological | 7.53 (6.75-8.41) | <0.001 | 4.48 (3.96-5.06) | <0.001 | 4.53 (4.00-5.12) | <0.001 | 4.52 (3.99-5.11) | <0.001 |
| Skin | 6.54 (5.54-7.72) | <0.001 | 6.62 (5.45-8.04) | <0.001 | 6.68 (5.50-8.11) | <0.001 | 6.69 (5.51-8.12) | <0.001 |
| Brain (malignant) | 24.56 (21.92-27.51) | <0.001 | 24.98 (21.98-28.39) | <0.001 | 25.04 (22.04-28.46) | <0.001 | 25.04 (22.03-28.45) | <0.001 |
| Other primary | 11.71 (10.05-13.64) | <0.001 | 10.22 (8.66-12.07) | <0.001 | 10.34 (8.75-12.20) | <0.001 | 10.29 (8.71-12.15) | <0.001 |
| Bone metastasis | 36.74 (33.18-40.70) | <0.001 | 12.92 (11.45-14.59) | <0.001 | 12.97 (11.49-14.65) | <0.001 | 13.05 (11.56-14.73) | <0.001 |
| Brain metastasis | 44.89 (40.54-49.72) | <0.001 | 21.42 (18.88-24.29) | <0.001 | 21.70 (19.14-24.61) | <0.001 | 21.78 (19.21-24.70) | <0.001 |
| Other metastasis | 24.58 (22.19-27.22) | <0.001 | 13.06 (11.60-14.71) | <0.001 | 13.18 (11.70-14.84) | <0.001 | 13.28 (11.80-14.96) | <0.001 |
| Treatment course intent | | | | | | | | |
| Curative | Ref | -- | Ref | -- | Ref | -- | Ref | -- |
| Oligomet. ablation | 4.18 (3.96-4.42) | <0.001 | 1.83 (1.69-1.98) | <0.001 | 1.83 (1.69-1.98) | <0.001 | 1.83 (1.69-1.98) | <0.001 |
| Palliative | 8.76 (8.44-9.09) | <0.001 | 3.42 (3.23-3.62) | <0.001 | 3.42 (3.23-3.62) | <0.001 | 3.44 (3.25-3.64) | <0.001 |
| Year treated | | | | | | | | |
| 2008-2015 | Ref | -- | Ref | -- | Ref | -- | Ref | -- |
| 2016-2023 | 0.74 (0.71-0.76) | <0.001 | 0.82 (0.79-0.86) | <0.001 | 0.82 (0.79-0.85) | <0.001 | 0.82 (0.79-0.86) | <0.001 |
| Radiation technique[e] | | | | | | | | |
| 3D | Ref | -- | Ref | -- | Ref | -- | Ref | -- |
| IMRT | 0.69 (0.66-0.72) | <0.001 | 0.75 (0.71-0.80) | <0.001 | 0.75 (0.71-0.80) | <0.001 | 0.75 (0.71-0.80) | <0.001 |
| SBRT | 1.08 (1.00-1.16) | 0.061 | 0.55 (0.51-0.60) | <0.001 | 0.55 (0.51-0.60) | <0.001 | 0.55 (0.51-0.60) | <0.001 |
| SRS | 2.57 (2.43-2.73) | <0.001 | 0.71 (0.65-0.77) | <0.001 | 0.71 (0.65-0.77) | <0.001 | 0.71 (0.66-0.78) | <0.001 |
| Brachytherapy | 0.20 (0.17-0.24) | <0.001 | 0.32 (0.26-0.39) | <0.001 | 0.32 (0.26-0.39) | <0.001 | 0.32 (0.26-0.39) | <0.001 |
| Electrons | 0.56 (0.49-0.64) | <0.001 | 0.61 (0.53-0.71) | <0.001 | 0.61 (0.53-0.71) | <0.001 | 0.61 (0.53-0.71) | <0.001 |
| Adaptive | 0.72 (0.60-0.86) | <0.001 | 0.55 (0.45-0.66) | <0.001 | 0.55 (0.45-0.66) | <0.001 | 0.55 (0.45-0.66) | <0.001 |

Abbreviations: FAD, FAHR-FaceAge deviation (FAHR-FaceAge minus chronological age); HR, hazard ratio; CI, confidence interval; Ref, reference; DCIS, ductal carcinoma in situ; oligomet., oligometastasis; 3D, 3-dimensional conformal; IMRT, intensity modulated radiotherapy; SBRT, stereotactic radiotherapy; SRS, stereotactic radiosurgery.
[a]n=38,211 with 13,769 events (64.0% censored), unless otherwise specified
[b]n=34,791 with 12,801 events. (63.2% censored)
[c]n=37,223

[d]Excluded from multivariate analysis given significant confounding with cancer site variable
[e]n-35,757

**Table 2. Univariate and multivariate Cox regression for overall survival – FAHR-FaceSurvival**

| Cox regression | Univariate[a] | | Multivariate model 1[b] (FAHR-FaceSurvival continuous) | | Multivariate model 2[b] (FAHR-FaceSurvival Quartiles) | |
|---|---|---|---|---|---|---|
| **Variables** | HR (95% CI) | P value | HR (95% CI) | P value | HR (95% CI) | P value |
| Age (decade) | 1.15 (1.11-1.19) | <0.001 | 1.08 (1.04-1.12) | <0.001 | 1.08 (1.04-1.13) | <0.001 |
| FAHR-FaceSurvival | | | | | | |
| [1]Continuous (0.1) | 1.32 (1.30-1.35) | <0.001 | 1.17 (1.15-1.20) | <0.001 | -- | -- |
| [2]Categorized | | | | | | |
| <0.25 | Ref | -- | -- | -- | Ref | -- |
| 0.25-0.49 | 2.30 (1.99-2.65) | <0.001 | -- | -- | 1.58 (1.36-1.84) | <0.001 |
| 0.5-0.74 | 4.29 (3.71-4.96) | <0.001 | -- | -- | 2.24 (1.91-2.30) | <0.001 |
| ≥0.75 | 8.73 (7.41-10.29) | <0.001 | -- | -- | 3.22 (2.69-3.85) | <0.001 |
| Sex, male vs female | 1.24 (1.11-1.37) | <0.001 | 1.03 (0.91-1.15) | 0.677 | 1.03 (0.92-1.16) | 0.643 |
| Race/ethnicity[c] | | | | | | |
| White | Ref | -- | Ref | -- | Ref | -- |
| Black | 0.95 (0.73-1.23) | 0.693 | 1.01 (0.78-1.33) | 0.918 | 0.98 (0.75-1.29) | 0.908 |
| Asian | 0.79 (0.56-1.12) | 0.183 | 0.72 (0.51-1.02) | 0.067 | 0.73 (0.52-1.04) | 0.080 |
| Hispanic | 0.88 (0.57-1.37) | 0.572 | 0.67 (0.42-1.07) | 0.096 | 0.65 (0.41-1.04) | 0.072 |
| Other | 1.42 (1.06-1.90) | 0.019 | 1.38 (1.02-1.85) | 0.036 | 1.39 (1.03-1.87) | 0.031 |
| Cancer risk group[d] | | | | | | |
| Non-metastatic cancer | Ref | -- | -- | -- | -- | -- |
| Benign/DCIS | 0.23 (0.12-0.43) | <0.001 | -- | -- | -- | -- |
| Metastatic cancer | 5.42 (4.86-6.03) | <0.001 | -- | -- | -- | -- |
| Cancer site | | | | | | |
| Breast | Ref | -- | Ref | -- | Ref | -- |
| Benign | 1.84 (0.74-4.62) | 0.192 | 1.21 (0.47-3.07) | 0.695 | 1.37 (0.54-3.50) | 0.506 |
| Prostate | 1.43 (0.91-2.26) | 0.124 | 1.68 (1.00-2.84) | 0.052 | 1.75 (1.04-2.96) | 0.036 |
| Lung | 12.79 (9.22-17.75) | <0.001 | 9.85 (6.78-14.32) | <0.001 | 10.26 (7.05-14.92) | <0.001 |
| Gastrointestinal | 8.80 (6.19-12.50) | <0.001 | 7.28 (4.96-10.69) | <0.001 | 7.66 (5.22-11.24) | <0.001 |
| Gynecological | 4.26 (2.78-6.51) | <0.001 | 3.66 (2.26-5.94) | <0.001 | 3.78 (2.33-6.14) | <0.001 |
| Head and neck | 4.54 (3.10-6.65) | <0.001 | 4.47 (2.89-6.93) | <0.001 | 4.67 (3.01-7.23) | <0.001 |
| Sarcoma | 4.26 (2.78-6.51) | <0.001 | 7.18 (4.34-11.87) | <0.001 | 7.47 (4.51-12.35) | <0.001 |
| Hematological | 7.31 (5.27-10.14) | <0.001 | 3.41 (2.36-4.93) | <0.001 | 3.51 (2.43-5.08) | <0.001 |
| Skin | 4.11 (2.37-7.13) | <0.001 | 3.27 (1.66-6.43) | <0.001 | 3.42 (1.74-6.73) | <0.001 |
| Brain (malignant) | 21.87 (15.53-30.80) | <0.001 | 16.96 (11.42-25.18) | <0.001 | 17.36 (11.69-25.80) | <0.001 |
| Other primary | 12.04 (7.54-19.22) | <0.001 | 7.47 (4.45-12.53) | <0.001 | 8.02 (4.78-13.47) | <0.001 |
| Bone metastasis | 34.23 (25.14-46.62) | <0.001 | 8.85 (6.08-12.88) | <0.001 | 9.07 (6.22-13.22) | <0.001 |
| Brain metastasis | 35.48 (26.18-48.08) | <0.001 | 10.83 (7.32-16.01) | <0.001 | 11.27 (7.61-16.68) | <0.001 |
| Other metastasis | 19.06 (14.02-25.91) | <0.001 | 7.96 (5.51-11.49) | <0.001 | 8.24 (5.70-11.90) | <0.001 |
| Treatment course intent | | | | | | |
| Curative | Ref | -- | Ref | -- | Ref | -- |
| Oligomet. ablation | 3.99 (3.34-4.77) | <0.001 | 1.62 (1.25-2.10) | <0.001 | 1.60 (1.23-2.07) | <0.001 |
| Palliative | 8.62 (7.68-9.66) | <0.001 | 3.38 (2.82-4.04) | <0.001 | 3.42 (2.86-4.10) | <0.001 |
| Year treated | | | | | | |
| 2008-2015 | Ref | -- | Ref | -- | Ref | -- |
| 2016-2023 | 0.73 (0.66-0.82) | <0.001 | 0.79 (0.70-0.90) | <0.001 | 0.81 (0.71-0.91) | <0.001 |
| Radiation technique[e] | | | | | | |
| 3D | Ref | -- | Ref | -- | Ref | -- |
| IMRT | 0.63 (0.55-0.72) | <0.001 | 0.67 (0.56-0.80) | <0.001 | 0.66 (0.55-0.79) | <0.001 |
| SBRT | 0.99 (0.76-1.29) | 0.946 | 0.58 (0.43-0.78) | <0.001 | 0.58 (0.43-0.77) | <0.001 |
| SRS | 2.48 (2.08-2.95) | <0.001 | 0.86 (0.66-1.12) | 0.266 | 0.87 (0.67-1.13) | 0.302 |
| Brachytherapy | 0.24 (0.14-0.41) | <0.001 | 0.49 (0.27-0.88) | 0.017 | 0.48 (0.27-0.87) | 0.016 |
| Electrons | 0.38 (0.25-0.60) | <0.001 | 0.62 (0.37-1.04) | 0.070 | 0.60 (0.36-1.00) | 0.049 |
| Adaptive | 0.46 (0.24-0.85) | 0.014 | 0.46 (0.24-0.88) | 0.018 | 0.44 (0.23-0.85) | 0.014 |

Abbreviations: cont, continuous; cat, categorized; HR, hazard ratio; CI, confidence interval; Ref, reference; DCIS, ductal carcinoma in situ; oligomet., oligometastasis; 3D, 3-dimensional conformal; IMRT, intensity modulated radiotherapy; SBRT, stereotactic radiotherapy; SRS, stereotactic radiosurgery.
[a]n=3,822 with 1,412 events (63.1% censored), unless otherwise specified
[b]n=3,480 with 1,318 events (62.1% censored)
[c]n=3,720
[d]Excluded from multivariate analysis given significant confounding with cancer site variable
[e]n=3,578

**Supplementary Materials**

**FAHR-Face (Foundation Model) Training Dataset**

The foundation model was trained on the WebFace42M[1] dataset, a cleaned subset of the WebFace260M dataset[1]. WebFace42M[1] consists of over 40 million high-quality facial images spanning 2 million unique identities, ensuring a diverse and comprehensive training set for machine learning tasks involving faces.

It contains individuals from over 200 distinct countries/regions and more than 500 professions, providing broad diversity in nationality and background. Analysis of the cleaned WebFace42M[1] shows it spans an extensive range of poses (yaw angles from -90 to 90 degrees) and includes most major races worldwide, such as Caucasian, African, East Asian, South Asian, and more[1]. The WebFace42M[1] dataset covers a wide range of ages, with dates of birth going back to 1846, ensuring substantial age diversity in the face images.

The dataset's diversity extends to image quality and lighting conditions. The images were collected from various sources across different time periods and scenarios (both controlled and unconstrained), so they represent a wide spectrum of photo qualities and lighting situations. This variety in image characteristics helps in training models that can generalize well to real-world applications where lighting and image quality may vary significantly.

The foundation model was trained on exactly 40,615,808 face images from approximately 2 million individuals with a size of 112x112 pixels. Besides the processing steps taken during training, no further preprocessing was performed on the images from WebFace42M[1]. The cleaning process used to create WebFace42M[1] from the larger WebFace260M[1] dataset likely removed many low-quality images, ensuring a baseline level of image quality while maintaining diversity in lighting and capture conditions.

**FAHR-Face (Foundation Model) Architecture and Training**

The foundation model is based on the Vision Transformer (ViT) architecture[2], specifically the base-sized model pre-trained using the Masked Autoencoder method. The "facebook/vit-mae-base[3]" model was released in the Facebook AI repository[3].

The ViT architecture is a transformer encoder model that processes images as a sequence of fixed-size patches. In this implementation, a patch size of 16x16 pixels is used. During pre-training with the Masked Autoencoder, a high portion (75%) of the image patches are randomly masked. The encoder encodes the visible patches, and a learnable mask token is added at the positions of the masked patches. The decoder then reconstructs the raw pixel values for the masked positions based on the encoded visible patches and mask tokens.

Several adjustments were made to adapt the "facebook/vit-mae-base[3]" model for this use case. First, the input images were resized to 112x112 pixels to match the resolution of the face photograph dataset, which altered the number of patches compared to the original 224x224 configuration. The positional embeddings for both the encoder and decoder were interpolated to handle the new patch arrangement while keeping the patch size at 16x16 pixels. The model was then fine-tuned on the WebFace42M[1] dataset using the AdamW optimizer with a learning rate of 1.5e-5, weight decay of 0.05, beta values of (0.9, 0.9999), and a batch size of 1024 over 23 epochs.

The fine-tuning process leverages transfer learning from the pre-trained weights of the "facebook/vit-mae-base[3]" model, allowing our foundation model to learn rich representations specific to facial analysis tasks. The model captures meaningful facial features and structures by utilizing the Masked Autoencoder objective during pre-training, providing a strong basis for subsequent tasks. Training, tuning, and testing were done on a workstation equipped with an AMD Ryzen Threadripper PRO 5975WX CPU, NVIDIA RTX A6000 GPU, and 252 GB RAM.

**FAHR-Face (Foundation Model) Evaluation**

The reconstruction ability of the trained foundation model is demonstrated in **Extended Data Figure 5**, which shows that the model can effectively capture and reconstruct critical facial features, structures, and details across a diverse range of ages, and races/ethnicities. Synthetic facial photographs (n = 22) were generated using Midjourney. The foundation model successfully reconstructed 75% of randomly masked regions in each photograph. Notably, the model employs an asymmetric encoder-decoder architecture with a purposely small encoder and larger decoder; this design ensures that the encoder learns compact and meaningful representations, with reconstruction serving as a training mechanism to guide representation learning rather than being the primary objective.

**FAHR-FaceAge (Age Estimation Model) Datasets**

The age estimation model combines eleven diverse facial age datasets to ensure a comprehensive and balanced training and testing set. The selected datasets include:

**IMDB-WIKI[4]:** The IMDB-WIKI dataset is a collection of 523,051 images annotated with age labels, collected from 100,000 celebrity profiles on IMDb and Wikipedia.

**KANFace[5]:** The KANFace dataset contains 41,036 images of 1,045 subjects captured in unconstrained, real-world conditions, with manual annotations for age.

**FG-NET[6]:** The FGNet dataset comprises 1002 face images from 82 individuals captured at different ages.

**CACD[7]:** The Cross-Age Celebrity Dataset consists of 163,446 face images of 2,000 celebrities collected from the Internet. The dataset was compiled using a combination of the celebrity's name and year (from 2004 to 2013) as search keywords, allowing the images to be labeled with an approximate age for each celebrity based on the year the photo was taken and their known birth year.

**AFAD[8]:** The Asian Face Age Dataset is a large-scale face image dataset. It consists of 164,432 facial images captured from Asian individuals and corresponding age and sex annotations. The AFAD was collected to address the lack of an extensive dataset for evaluating age estimation performance on Asian faces.

**MegaAge[9]:** The MegaAge dataset consists of 41,941 face images collected from the Internet with age labels. The dataset was compiled using an approach that transfers reliable age annotations from the FG-NET database to unlabeled images via human-in-the-loop comparisons.

**MORPH[10]:** The MORPH dataset contains 55,134 images with age labels of 13,617 individuals. The dataset was collected from public records.

**LAGENDA[11]:** The Layer Age and Gender Dataset consists of 67,159 images with 83,144 annotated face images sourced from the Open Images Dataset, covering various real-world scenes and domains with minimal celebrity data.

**UTK[12]:** The UTKFace dataset is a face image dataset that spans a wide age range. It consists of 24,108 face images with annotations for age. The dataset includes unconstrained faces with pose, expression, illumination, occlusion, and resolution variations.

**AgeDB[13]:** The AgeDB dataset is manually collected, comprising 16,488 in-the-wild images of well-known individuals like actors, scientists, and politicians, each annotated with age a total of 568 distinct subjects.

**APPA-REAL[14]:** The APPA-REAL dataset was created by combining community inputs through a web application and data from the AgeGuess platform, resulting in 7,591 images with age labels. The dataset consists primarily of photos contributed by the general public rather than celebrities.

The photos from IMDB-WIKI[4], FG-NET[6], CACD[7], MegaAge[9], MORPH[10], UTK[12], AgeDB[13], APPA-REAL[14] were cropped and aligned with RetinaFace[15]; only photos where exactly one face was detected were used. For LAGENDA[11], KANFace[5], and AFAD[8], the already cropped images provided in these datasets were used.

The dataset was divided into three categories: Pre-training, which includes IMDB-WIKI[4], KANFace[5], FG-NET[6], CACD[7], AFAD[8], MegaAge[9], MORPH[10], LAGENDA[11]; Fine-tuning, comprised of UTK[12], AgeDB[13] (chosen specifically for their more balanced age distribution and representation of older subjects); Evaluation, focused on APPA-REAL[14]( (chosen specifically for its more balanced age distribution, representation of older subjects, and primary contribution from the general public, making overlap with other datasets unlikely). **Extended Data Figure 6** illustrates the dataset filtering and preprocessing pipeline applied to each dataset before combining them into the final training, fine-tuning, and evaluation sets. The pipeline begins with the raw datasets, each containing a unique set of face photos with corresponding age labels. After the filtering and preprocessing steps, the age distribution of each dataset is analyzed and visualized in **Extended Data Figure 6.**

The multi-dataset approach in the training data promotes high generalization, allowing the model to perform well on unseen faces and adapt to various real-world scenarios.

**FAHR-FaceAge Model Adjustments and Training**

The FAHR-FaceAge was built upon our foundation model, a Vision Transformer (ViT) pre-trained using the Masked Autoencoder method. The model architecture was modified to include a regression head for age estimation, consisting of a single linear layer that maps the hidden state of the first token (CLS token) to a scalar age value.

Data augmentation techniques were applied during training to enhance the model's robustness and generalization ability. The augmentation steps included Random Resized Crop with size (112, 112) and scale (0.8, 1.0), Random Horizontal Flip with a probability of 0.5, Color Jitter with brightness 0.4, contrast 0.4, saturation 0.4, hue 0.1, and probability of 0.8, Random Grayscale with probability 0.2, Random Rotation with 10 degrees, Gaussian Blur with kernel size (5, 5), sigma (0.1, 2.0), and probability of 0.5, Random Erasing with probability of 0.5, scale (0.02, 0.33), ratio (0.3, 3.3), and value 0, and Normalization with mean [0.485, 0.456, 0.406] and standard deviation [0.229, 0.224, 0.225].

The training process was divided into two stages: pre-training and fine-tuning. The model was trained on a combination of eight diverse facial age datasets described above for pre-training. These datasets collectively provided various face images with corresponding age labels. However, the age distribution in these datasets was imbalanced, with certain age groups (e.g., elderly individuals) underrepresented.

An oversampling strategy was applied during pre-training and fine-tuning to correct the imbalance and ensure the model learned from a more evenly distributed age range. The oversampling factors for pre-training were determined based on the age range, with more aggressive oversampling applied to the underrepresented age groups. Images of individuals aged 0-50 years were not oversampled (factor = 1), while images of individuals in the 50-55 age range were oversampled by a factor of 2. The oversampling factor gradually increased for older age ranges, with the 95-116 age range having the highest oversampling factor of 20. This oversampling ensured that the model was exposed to sufficient samples from the underrepresented and rare age groups during pre-training.

During pre-training, the model was optimized using the AdamW optimizer with a base learning rate of 1e-5, weight decay of 0.05, beta values of (0.9, 0.999), and a layer-wise learning rate decay of 0.75. The model was trained for 30 epochs with a batch size of 64. The mean absolute error (MAE) was used as the loss function during pre-training.

The model was further fine-tuned on the UTK[12] and AgeDB[13] datasets. A different oversampling strategy was employed during fine-tuning to ensure a more even distribution of ages. The fine-tuning dataset was divided into age bins of 5 years, and each bin was balanced to a specified size of 200 samples. If a bin contained more than 200 samples, a random subset of 200 samples was selected. A bin with fewer than 200 samples was oversampled by randomly duplicating samples until the desired size was reached. This binning and balancing approach guaranteed that each age group had equal representation during fine-tuning. Both oversampling strategies, pre-training and fine-tuning, are shown in **Extended Data Figure 6.**

The fine-tuning stage utilized the AdamW optimizer with a reduced base learning rate of 1e-6, weight decay of 0.05, beta values of (0.9, 0.999), and a layer-wise learning rate decay of 0.65. The model was fine-tuned for 10 epochs with a batch size of 8, using the MAE loss as the optimization objective.

By employing these oversampling strategies, hyperparameter settings, and the custom learning rate schedule, FAHR-FaceAge was able to learn effectively from the imbalanced pre-training datasets and achieved improved performance on the fine-tuning datasets.

**FAHR-FaceAge Evaluation**

FAHR-FaceAge was evaluated on the public APPA-REAL[14] and Harvard RT patient datasets. Our model's age estimation performance in these two distinct datasets (real-world healthy individuals and cancer patients, respectively) was compared against several models representing a diverse range of state-of-the-art architectures and have been widely used in computer vision tasks, including VGG16[16], SENet50[17], DenseNet121[18], and four different versions of MobileNet[19] (V2-96, V2-224, V3-small, and V3-large) as implemented by Greco et al.[20]. These architectures were trained either through knowledge distillation (using VGGFace2[21] images with age labels generated by a teacher model to create VMAGE) or with IMDB-Wiki[4] dataset, and evaluated on the same cropped face photographs as FAHR-FaceAge. While FAHR-FaceAge uses an input size of 112×112 pixels for models that support higher resolution inputs (e.g., 224×224), we resized the images accordingly to leverage their full capabilities.

For the APPA-REAL dataset, the MAE comparison showed that FAHR-FaceAge has the lowest MAE at 5.1, followed by SENet50 VMAGE at 5.5, MobileNet V3-large VMAGE at 5.6, MobileNet224 VMAGE at 5.9, MobileNet V3-small VMAGE at 5.9, MobileNet96 VMAGE at 6.1, DenseNet121 VMAG at 6.4, DenseNet IMDB-WIKI at 7.9, MobileNet V3-large IMDB-WIKI at 8.0, MobileNet V3-small IMDB-WIKI at 8.0, MobileNet224 IMDB-WIKI at 8.2, MobileNet96 IMDB-WIKI at 8.6, SENet50 IMDB-WIKI at 8.8, FaceAge at 12.6, VGG16 IMDB-WIKI at 16.5, and VGG16 VMAGE at 63.7.

For the Harvard RT patient dataset, the MAE comparison showed that FaceAge has the lowest MAE at 7.2, followed by FAHR-FaceAge at 7.5, MobileNet V3-large IMDB-WIKI at 9.4, SENet50 IMDB-WIKI at 10.6, SENet50 VMAGE at 11.3, MobileNet V3-large VMAGE at 11.7, MobileNet224 VMAGE at 11.7, DenseNet IMDB-WIKI at 11.8, MobileNet V3-small IMDB-WIKI at 12.1, MobileNet V3-small VMAGE at 12.2, Mobilenet96 VMAGE at 12.5, DenseNet121 VMAGE at 13.2, MobileNet224 IMDB-WIKI at 13.3, MobileNet96 IMDB-WIKI at 14.5, VGG16 IMDB-WIKI at 27.1, and VGG16 VMAGE at 30.5.

In addition to reporting the MAE across all samples, we evaluated each model using the average bin-wise MAE: for each 5-year age bin (starting at age 0), we calculated the MAE and then averaged these values across all bins. This metric better captures model robustness to age bias and reflects overall accuracy across the lifespan, penalizing models that perform poorly in specific age groups.

For the APPA-REAL dataset, the Average bin-wise MAE comparison showed that FAHR-FaceAge has the lowest value at 5.7, followed by SENet50 VMAGE at 9.7, MobileNet V3-large VMAGE at 10.0, MobileNet224 VMAGE at 10.2, MobileNet V3-small VMAGE at 10.4, MobileNet96 VMAGE at 10.7, DenseNet121 VMAG at 11.0, MobileNet V3-large IMDB-WIKI at 11.2, MobileNet224 IMDB-WIKI at 11.7, FaceAge at 11.8, DenseNet IMDB-WIKI at 12.1, SENet50 IMDB-WIKI at 12.3, MobileNet V3-small IMDB-WIKI at 12.4, MobileNet96 IMDB-WIKI at 12.7, VGG16 IMDB-WIKI at 21.6, and VGG16 VMAGE at 45.7.

For the Harvard RT patient dataset, the Average bin-wise MAE comparison showed that FAHR-FaceAge has the lowest value at 6.1, followed by SENet50 VMAGE at 10.5, MobileNet V3-large VMAGE at 10.8, MobileNet V3-large IMDB-WIKI at 11.0, MobileNet224 VMAGE at 11.0, FaceAge at 11.3, MobileNet V3-small VMAGE at 11.3, MobileNet96 VMAGE at 11.8, DenseNet121 VMAGE at 11.9, DenseNet IMDB-WIKI at 12.3, MobileNet V3-small IMDB-WIKI at 12.7, SENet50 IMDB-WIKI at 13.0, MobileNet224 IMDB-WIKI at 13.2, MobileNet96 IMDB-WIKI at 14.5, VGG16 IMDB-WIKI at 22.4, and VGG16 VMAGE at 43.8.

The 5 best-performing models for each dataset are visualized in **Extended Data Figure 1**.

**Benchmark on Fine-tuned Models**

To ensure a fair comparison, we fine-tuned each competing architecture using the same training configuration as FAHR-FaceAge. Most training parameters, including batch size, number of epochs, and dataset composition (oversampled UTK and AgeDB), remained identical; only the learning rate was adjusted, and augmentation pipeline was changed. Fine-tuning these pretrained models proved non-trivial, as they had initially learned from highly age-biased datasets, primarily consisting of young adult faces. Expanding the target age range to span the full human lifespan, from young children to elderly adults, increases the complexity of the age-estimation task, inherently leading to higher MAE values compared to narrower age-range benchmarks. Pilot experiments indicated that higher learning rates quickly disrupted the pretrained age-specific features, resulting in poor convergence and generalization. Therefore, we carefully selected a conservative learning rate of 1e-7 to ensure stable training and optimal performance across the entire age spectrum. During training, each image undergoes the following augmentations: random rotation, random affine skew, random crop (by jittering the region of interest), random brightness and contrast adjustment, and random horizontal flip (p=0.5). These augmentations are the default augmentation from Greco et al.[20] and applied to every image in training.

Despite this harmonized fine-tuning process, FAHR-FaceAge consistently outperformed the benchmarked models across both evaluation datasets in terms of accuracy and bias, demonstrating the advantage of starting from a large-scale, self-supervised foundation model.

For the APPA-REAL dataset, the MAE comparison showed that FAHR-FaceAge has the lowest MAE at 5.1, followed by MobileNet224 VMAGE at 6.5, SENet50 VMAGE at 6.9, MobileNet V3 large VMAGE at 7.0, MobileNet96 VMAGE at 7.1, DenseNet IMDB-WIKI at 7.9, MobileNet96 IMDB-WIKI at 8.6, MobileNet224 IMDB-WIKI at 9.0, SENet50 IMDB-WIKI at 9.0, MobileNet V3 large IMDB-WIKI at 9.2, MobileNet V3 small IMDB-WIKI at 9.4, DenseNet121 VMAG at 10.0, FaceAge at 12.6, VGG16 IMDB-WIKI at 14.8, VGG16 VMAGE at 15.2, and MobileNet V3 small VMAGE at 30.3.

For the Harvard RT patient dataset, the MAE comparison showed that FaceAge has the lowest MAE at 7.2, followed by FAHR-FaceAge at 7.5, SENet50 IMDB-WIKI at 8.7, DenseNet IMDB-WIKI at 10.9, SENet50 VMAGE at 12.6, MobileNet V3 small VMAGE at 13.5, MobileNet V3 large IMDB-WIKI at 14.3, MobileNet V3 large VMAGE at 14.4, DenseNet121 VMAGE at 14.5, MobileNet224 VMAGE at 14.9, MobileNet224 IMDB-WIKI at 15.7, Mobilenet96 VMAGE at 16.7, MobileNet96 IMDB-WIKI at 16.9, MobileNet V3 small IMDB-WIKI at 17.8, VGG16 IMDB-WIKI at 18.1, and VGG16 VMAGE at 18.4.

For the APPA-REAL dataset, the Average bin-wise MAE comparison showed that FAHR-FaceAge has the lowest value at 5.7, followed by SENet50 VMAGE at 10.3, SENet50 IMDB-WIKI at 10.7, FaceAge at 11.8, DenseNet IMDB-WIKI at 12.1, MobileNet224 VMAGE at 12.3, MobileNet V3-large VMAGE at 12.4, DenseNet121 VMAG at 13.1, MobileNet224 IMDB-WIKI at 13.5, MobileNet96 IMDB-WIKI at 13.7, MobileNet96 VMAGE at 14.0, MobileNet V3-large IMDB-WIKI at 14.8, VGG16 VMAGE at 14.8, MobileNet V3-small IMDB-WIKI at 16.7, VGG16 IMDB-WIKI at 16.8, and MobileNet V3-small VMAGE at 48.4.

For the Harvard RT patient dataset, the Average bin-wise MAE comparison showed that FAHR-FaceAge has the lowest value at 6.1, followed by SENet50 VMAGE at 10.7, FaceAge at 11.3, SENet50 IMDB-WIKI at 11.3, DenseNet IMDB-WIKI at 12.0, MobileNet V3-small VMAGE at 12.3, MobileNet V3-large VMAGE at 12.7, MobileNet224 VMAGE at 12.8, DenseNet121 VMAGE at 13.4, VGG16 VMAGE at 14.2, MobileNet96 VMAGE at 14.4, VGG16 IMDB-WIKI at 14.5, MobileNet V3-large IMDB-WIKI at 15.0, MobileNet224 IMDB-WIKI at 15.3, MobileNet96 IMDB-WIKI at 15.4, and MobileNet V3-small IMDB-WIKI at 17.5.

The 5 best-performing models for each dataset are visualized in **Extended Data Figure 1**.

**FAHR-FaceAge Robustness Evaluation Datasets**

We evaluated our model's robustness across three publicly available datasets that capture different types of variations in facial appearance or facial photograph acquisition:

**HDA Plastic Surgery Database**[22]: Contains 638 image pairs (540 female, 98 male) showing faces before and after five common plastic surgery procedures: eyebrow correction, eyelid correction, facelift, facial bones correction, and nose correction. All images were collected from multiple web sources. Each image pair shows the same individual before and after a single type of surgical procedure.

**Makeup Dataset**[23]: Consists of 26 paired images of female subjects with and without makeup application. The dataset includes both subtle and dramatic makeup changes. All images were collected under controlled conditions with consistent lighting and frontal pose.

**FEI Face Database**[24]: A controlled face pose dataset containing 200 subjects (100 male, 100 female) with 14 different pose angles per subject. From the original 2800 images, we used a subset focusing on three key variations: frontal face comparisons with expression changes (neutral and smiling), illumination variations, and pose variations (22.5° left/right vs. frontal).

All images from the three datasets were pre-processed using RetinaFace for face detection and alignment to ensure consistent facial presentation and image quality.

**FAHR-FaceAge Robustness Evaluation**

FAHR-FaceAge's robustness to facial alterations and viewing conditions was evaluated using paired images processed with RetinaFace for face detection and alignment. For each image pair, the mean difference and mean absolute difference in age estimation were calculated by subtracting the first condition from the second. Statistical significance was tested using paired Wilcoxon tests.

For plastic surgery evaluation (HDA database[22]), the results were: eyebrow correction (mean difference=0.29±6.30 years, mean absolute difference=4.62 years, n=126), eyelid correction (mean difference=-0.45±5.85 years, mean absolute difference=4.26 years, n=130), facelift (mean difference=-0.40±5.64 years, mean absolute difference=4.53 years, n=96), facial bones correction (mean difference=1.03±6.16 years, mean absolute difference=4.63 years, n=105), and nose correction (mean difference=1.11±4.21 years, mean absolute difference=3.13 years, n=174). For makeup analysis[23], we found a mean difference=1.09±6.31 years and a mean absolute difference=4.37 years (n=26). The FEI database[24] showed viewing angle differences (left: mean difference=0.25±1.49 years, mean absolute difference=1.12 years; right: mean difference=0.55±1.65 years, mean absolute difference=1.17 years; n=200), expression changes (mean difference=0.79±3.68 years, mean absolute difference=2.19 years, n=200), and illumination variations (mean difference=0.55±3.77 years, mean absolute difference=2.35 years, n=181).

**Random initialization baseline**

To isolate the contribution of foundation-model pre-training, we trained a ViT with random weight initialization. Architecture, optimizer hyper-parameters, data splits, augmentations, batch size, and number of epochs were kept identical to FAHR-FaceAge; the sole change was disabling layer-wise learning-rate decay.

The baseline ViT trained from random initialization reached 6.92 years MAE (ME 1.29 years) on APPA-Real and 9.11 years MAE (ME 0.25 years) on the Harvard RT dataset.

**FAHR-FaceSurvival (Risk Estimation Model) Dataset Considerations**

The Harvard RT dataset (described in further data below) was randomly split into two parts: 90% for training and 10% for testing. During the training process, 5% of the training data was used for validation. The test dataset is not used for validation, and the model never sees this test dataset during training or validation. The Harvard RT dataset also includes vital status, the last date of follow-up if the patient is alive, and the date of death for survival analysis.

**FAHR-FaceSurvival (Risk Estimation Model) Model Integration and Architecture**

A custom dataset loader was implemented to handle the dataset loading and preprocessing. It read the survival data from a CSV file and loaded the corresponding face photos from a specified folder. The dataset returned each sample's face photo, survival time, and outcome (death or censored) status.

Data augmentation techniques were applied to the training set to improve the model's robustness and generalization ability. The validation set underwent resizing and normalization. The specific augmentations and preprocessing steps included Random Resized Crop with size (112, 112) and scale (0.9, 1.0), Random Horizontal Flip with a probability of 0.5, Color Jitter with brightness 0.2, contrast 0.2, saturation 0.2, hue 0.1, and a probability of 0.8, Random Rotation with 5 degrees, and Normalization with mean [0.485, 0.456, 0.406] and standard deviation [0.229, 0.224, 0.225].

The FAHR-Face was initialized with the pre-trained weights. The survival head was a linear layer that mapped the hidden representation of the first token (CLS token) to a scalar FAHR-FaceSurvival. The model architecture allowed for fine-tuning the pre-trained weights while learning the task-specific risk estimation.

The training process utilized a custom ranking loss function designed to optimize the ranking of the FAHR-FaceSurvival based on survival times. To achieve this, the loss function first identified pairs of patients within each training batch, where each pair consisted of one patient who experienced the event of interest (death), thus representing a complete observation, and another patient whose survival time was observed to be longer. All possible pairs in each batch were used. The loss function encouraged the model to assign a higher FAHR-FaceSurvival to patients with shorter survival times compared to those with longer survival times. The loss function also incorporated a smoothed regularization term to encourage smooth FAHR-FaceSurvival predictions. The smoothed regularization term in the ranking loss function promoted the model to produce FAHR-FaceSurvival predictions that were smooth and consistent across similar patients by penalizing large differences between the FAHR-FaceSurvival of adjacent patients in the sorted order, promoting a more stable and coherent risk prediction.

During training, the model was optimized using the AdamW optimizer with a base learning rate of 1e-5, weight decay of 0.05, beta values of (0.9, 0.999), layer-wise learning rate decay of 0.75, batch size of 32, and smoothed regularization strength of 1e-4. The model was trained for 10 epochs, and checkpoints were saved after each epoch. The model's performance was evaluated using a custom concordance index (C-index) function, which measured the model's ability to rank the FAHR-FaceSurvival correctly. The C-index was computed separately for each epoch's training and validation sets. The training and validation losses were also monitored and reported.

**FAHR-FaceSurvival (Risk Estimation Model) Evaluation**

FAHR-FaceSurvival was evaluated on two datasets: the Harvard RT Test dataset and the Maastro RT dataset. FAHR-FaceSurvival was min-max scaled to a range between 0 and 1 for each dataset to ensure consistency and facilitate comparative analysis.

**Study Population Harvard RT Patient Dataset.**

The Harvard RT Dataset consists of 38,211 patients treated with radiation therapy for cancer or benign conditions between 2008 and 2023 at a tertiary Harvard-affiliated academic medical center and four affiliated community network clinical sites. Diagnosis associated with the radiation therapy course for each patient was obtained based on the international classification of diseases ICD-9 or ICD-10 codes. ICD codes were categorized into diagnosis and cancer site groups used in the study analysis, as detailed in **Extended Data Table 4**. Following the classification of diagnosis based on the ICD codes, available case descriptions summarizing the diagnosis, including stage and radiation plan, were used to categorize cases further. For example, if a patient had an ICD-10 code of C61 indicating prostate cancer, but the case description indicated "metastatic" prostate cancer, then the diagnosis was re-classified as metastatic disease. For oropharyngeal cancers, if the description indicated "HPV+ (human papillomavirus)" type, then the diagnosis was further specified as HPV+ oropharyngeal cancer. Benign diseases included keloids, heterotopic ossification, non-malignant brain tumors (pituitary adenomas, schwannomas, meningiomas), brain arteriovenous malformation, and cardiovascular disease (ventricular tachycardia, coronary artery disease).

From the initial 46,770 patients, 927 photos were excluded due to poor quality or substantial portions of the face being hidden (e.g., face masks, sunglasses), resulting in 45,843 remaining patients. The photos of these 45,843 patients were processed using the RetinaFace[25] face detection and alignment algorithm, retaining only those where exactly one face was detected, resulting in a dataset of 44,182 patients (1,661 patients excluded). Patients with missing follow-up information or cancer type were further excluded (5,971 patients), leading to a final cohort of 38,211 patients for analysis. Detailed characteristics of the Harvard RT Dataset are provided in **Extended Data Table 1**.

**External Validation - Maastro Dataset**

The Maastro Dataset consists of 4,892 patients with cancer diagnoses, prospectively collected and included in the Maastro Biobank (Maastricht, The Netherlands). These patients were treated with radiation therapy with curative or palliative intent between 2006 and 2019. The predominant primary malignancies were breast, colorectal, prostate, lung, and head and neck cancer. From an initial dataset of 6,835 patients, we excluded patients with missing face photographs, duplicate entries, and lack of follow-up information. A manual image quality assessment was performed on the remaining entries; patients were excluded due to poor image quality, leaving us with 5,498 patients. Patients with metastatic disease, those treated with palliative intent, or those diagnosed with ductal carcinoma in situ of the breast (DCIS) were subsequently removed from the dataset. This step reduced the cohort to 4,906 patients.

In the final analysis, 4,892 patients were included. This final selection was based on the use of a face detection model, RetinaFace[25] that only selected photos with precisely one person, ensuring consistency in FAHR-FaceAge and FAHR-FaceSurvival predictions. The dataset includes information on patient demographics, cancer type, treatment details, and survival outcomes. Detailed characteristics of the Maastro Dataset are provided in **Extended Data Table 1**.

**Random initialization baseline**

To isolate the contribution of foundation-model pre-training, we trained a ViT with random weight initialization. Architecture, optimizer hyper-parameters, data splits, augmentations, batch size, and number of epochs were kept identical to FAHR-FaceSurvival; the sole change was disabling layer-wise learning-rate decay.

**Interpretability, Independence, and Biological Basis of the Models – 2D and 3D Attention Maps**

Attention maps were utilized, generated from the trained models FAHR-FaceAge and FAHR-FaceSurvival, to reveal which regions of the face contributed most to each model's predictions and provide insights into the independence of the model's decision-making process despite a shared foundation model.

The maps were generated by preprocessing the images with standard transformations, such as resizing to 112x112 pixels and normalization. The average attention map was then computed across the entire Harvard RT Test datasets. Attention weights were extracted from the final model layer, averaged across all attention heads, and then interpolated to a higher resolution of 448x448 pixels using bilinear interpolation to enhance visualization quality.

For a spatially accurate visualization of the facial regions that contribute most to the model's predictions, three-dimensional attention maps were generated by mapping attention scores onto a 3D facial mesh. Facial images from the Harvard RT test dataset were preprocessed by resizing them to 112×112 pixels and normalizing each color channel using standard mean and standard deviation values. These preprocessed images were input into the trained FAHR-FaceAge and FAHR-FaceSurvival models to obtain attention maps. Attention weights were extracted from the final self-attention layer, focusing specifically on the attention between the class token and the image patches. These weights were averaged across all attention heads to produce a single attention map per image. Given the input image size and patch size, the resulting attention map had dimensions of 7×7, up sampled to 112×112 pixels using bilinear interpolation to match the original image resolution.

Facial landmarks were detected using the MediaPipe Face Mesh algorithm, providing 468 three-dimensional facial landmarks per image. The detected landmarks were scaled to align with the 112×112 pixel resolution of the attention maps for each photograph. A canonical 3D face model was a reference mesh for mapping the attention scores. The mesh resolution was enhanced by performing one iteration of triangle subdivision, subdividing each triangle into four smaller ones, and interpolating new vertices accordingly. Facial landmarks were updated to include positions for the new vertices through linear interpolation between the original landmarks.

For each image, the attention scores from the two-dimensional attention maps were mapped onto the subdivided 3D face mesh. The 3D mesh was projected onto the two-dimensional attention map using facial landmarks, mapping the 3D vertices to 2D coordinates on the attention map. This projection assumes a frontal camera perspective and does not account for facial rotations or tilts. For each triangle in the mesh, a polygon was defined using the projected 2D coordinates of its vertices, and a binary mask for the triangle's area on the attention map was created. The average attention value within each triangle's area was calculated by averaging the attention map values corresponding to the triangle's mask, providing an attention score for each triangle.

A comprehensive representation of each model's attention distribution was obtained by averaging the attention scores across all images in the dataset. This was done by summing the attention scores for each triangle across all images and then dividing the total by the number of images. The averaged attention scores were visualized by mapping them to colors using a standard colormap, with higher scores corresponding to warmer colors. Each triangle was assigned a color based on its attention score, and the colored 3D face mesh was saved in the OBJ file format, including vertex positions and per-face colors, allowing for interactive visualization in 3D rendering software.

Attention mapping onto a standardized 3D facial mesh across the dataset provided a more quantitative visualization of the facial regions that the models focus on when making predictions.

**Statistical Analysis**

Descriptive statistics were used to summarize baseline patient characteristics in the two patient cohorts. For survival analysis, the primary endpoint was overall survival (OS), calculated from the time the facial photograph was obtained for RT. Univariate Cox proportional hazard (PH) analysis was performed to assess the effect of FAHR-FaceAge or FAHR-FaceSurvival variables as well as clinical covariables including age, sex, race/ethnic group, diagnosis, treatment course intent, year treated, and radiation technique on mortality risk. A multivariable Cox PH model was generated incorporating FAHR-FaceAge or FAHR-FaceSurvival variables and all covariables with P-value <0.05 on univariate analyses. Limited key covariables were included and specified in the results table footnotes for subgroup analyses with fewer events. FAHR-FaceAge parameters assessed were FAHR-FaceAge as a continuous variable (per decade), and the difference between FAHR-FaceAge and chronological age (FAD) as a continuous variable (per decade), FAD ≥5 vs <5, and FAD ≤-5 vs >-5. FAHR-FaceSurvival parameters assessed were FAHR-FaceSurvival as a continuous variable (per 0.1 from zero to one), FAHR-FaceSurvival as quartiles (<0.25 vs. 0.25-0.49, 0.50-0.74 vs. ≥0.75), and FAHR-FaceSurvival as a binary split (<0.5 vs. ≥0.5).

Detailed results of the univariable and multivariable Cox regression analyses for FAHR-FaceAge and FAHR-FaceSurvival within various age, racial, and clinical subgroups in the Harvard RT dataset are provided in **Extended Data Table 2** and **Extended Data Table 3**, respectively, and for the Maastro dataset in **Figure 4** respectively.

Harrell's concordance index (C-index) was utilized to evaluate the predictive performance of the FAHR-FaceAge model in **Figure 2c** for the Harvard RT dataset and of the FAHR-FaceSurvival in **Figure 3g** for the Harvard RT Test dataset. Subgroup analyses evaluated the model's performance across different demographic and clinical groups, including age groups, racial and ethnic categories, and cancer types.

### Area Under the Curve (AUC) Analysis

The predictive performance of the FAHR-FaceAge was benchmarked against a prior deep learning-based facial age estimation model FaceAge[26], for the clinical application of survival prediction by using time-dependent receiver operating characteristic (ROC) curves and the area under the curve (AUC) metric in the Maastro RT dataset and Harvard RT dataset.

Data were extracted from Maastro dataset and Harvard RT dataset, focusing on the subset of patients older than 60 years because the prior FaceAge[26] model was only trained and validated for individuals over 60.

Time points of interest were set at 3 months, 6 months, 1 year, and 2 years after baseline. Time-dependent ROC curves were constructed using the Kaplan-Meier estimator method for each time point. The predictors evaluated included chronological age, FaceAge (age estimated by the FaceAge model[26]), and FAHR-FaceAge. The AUC was calculated for each predictor and time point to assess the model's discriminative ability at that specific time.

### Analysis of Independence of Imaging Biomarkers from the Two Models

To further quantitatively demonstrate the independence of FAHR-FaceAge and FAHR-FaceSurvival despite their shared foundation model, we analyzed their correlation in both development and external validation cohorts. In the Harvard RT Test dataset, FAD and FAHR-FaceSurvival showed a weak positive correlation (R = 0.218; **Extended Data Figure 4d**). This independence was further validated in the Maastro RT dataset, where FAD and FAHR-FaceSurvival maintained a low correlation (R = 0.160; **Figure 4e**), suggesting these measures capture distinct aspects of patient health even in an independent cohort.

The feature space independence was quantified by computing the cosine similarity between the final hidden state representations (768-dimensional class token embeddings) of both models for each patient image. Cosine similarity measures the cosine of the angle between two vectors, ranging from -1 (completely opposite) to 1 (exactly aligned), with 0 indicating orthogonality. The median similarity of -0.018 across all test patients indicates near-orthogonal feature spaces, suggesting the models learned to extract distinct and complementary features despite sharing the same foundation model architecture (**Extended Data Figure 4e**).

To assess the incremental prognostic value of our imaging biomarkers, we conducted a model comparison using AIC. Three Cox proportional hazards models were fitted: one including FAD as a continuous variable (per decade), one including FAHR-FaceSurvival as a continuous variable (per 0.1 unit increase), and a combined model incorporating both biomarkers, with adjustments for age (in decades), sex, and cancer site. AIC was calculated as AIC = -2 log(L) + 2k, where L is the likelihood of the model and k is the number of estimated parameters; lower AIC values denote a better trade-off between model fit and complexity.

All statistical analyses were performed using R Statistical Software v4.3.1 and SAS v9.4 (SAS Institute Inc). Statistical significance was defined as a two-sided P-value <0.05

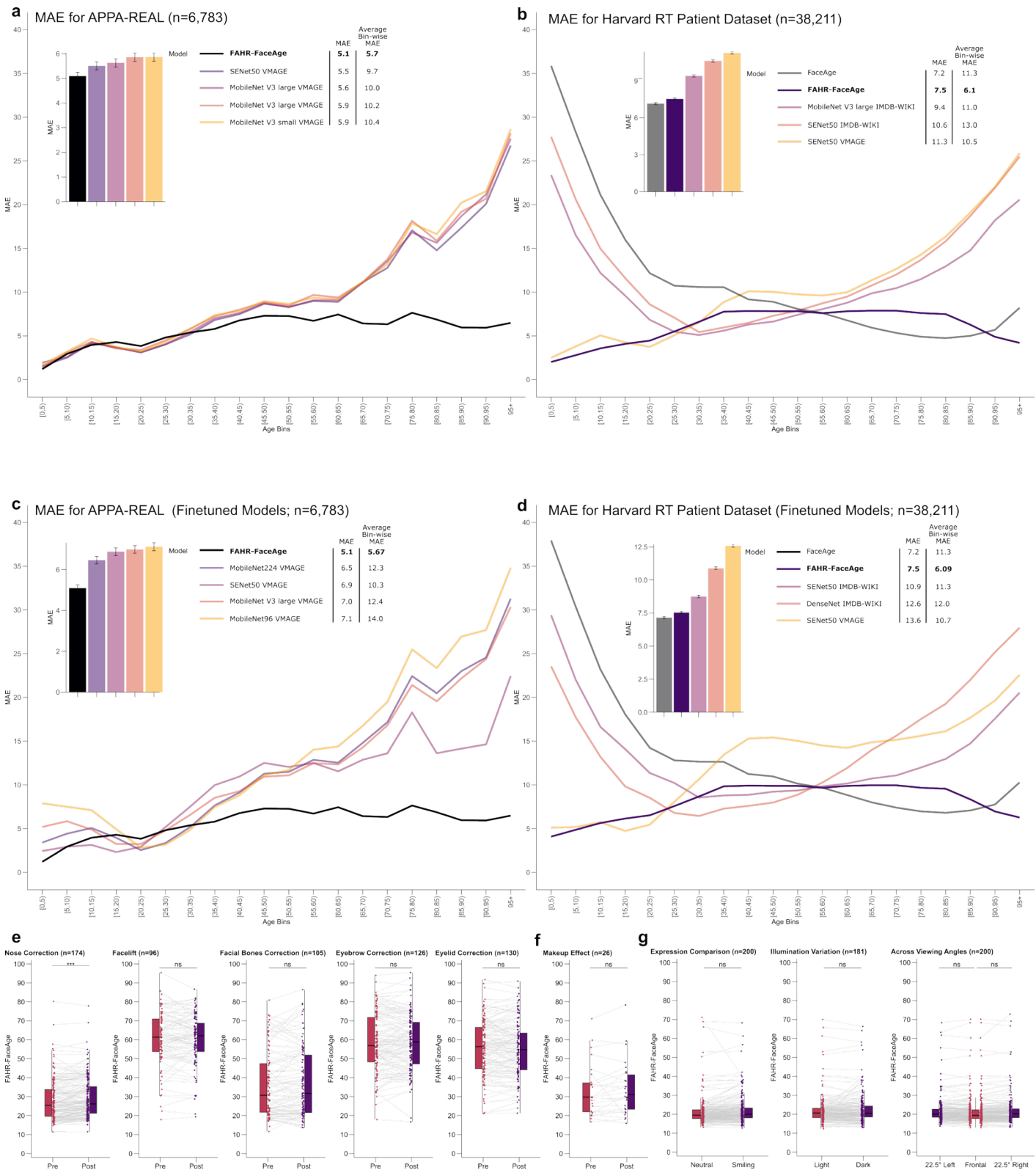

**Extended Data Figure 1: State-of-the-Art Comparison for Age Estimation Models and Robustness Test**

Each panel presents the performance of the top five age estimation models from face photographs, demonstrating the FAHR-FaceAge model's superior performance in both mean absolute error (MAE) and average bin-wise MAE across different datasets. Notably, for the Harvard RT patient dataset, FaceAge achieves a lower MAE due to its optimization for older patients (>60 years), which aligns well with the age distribution of the cancer cohort. However, FaceAge does not rank among the top five models for the APPA-REAL dataset, which is attributed to this dataset's younger age distribution. MAE is reported in years. Error bars represent 99 % confidence intervals.

**a,** MAE for the APPA-REAL dataset, shown as a bar plot. The line plot displays performance for each 5-year age bin for the top five models. The models included are FAHR-FaceAge, SENet50 VMAGE, MobileNet V3-large VMAGE, MobileNet224 VMAGE and MobileNet V3-small VMAGE.

**b,** MAE for the Harvard RT patient dataset, shown as a bar plot. The line plot displays performance for each 5-year age bin for the top five models. The models included are FaceAge, FAHR-FaceAge, MobileNet V3-large IMDB-WIKI, SENet50 IMDB-WIKI and SENet50 VMAGE.

**c,** MAE for the APPA-REAL dataset after fine-tuning on age-balanced dataset, shown as a bar plot. The line plot displays the performance for each 5-year age bin for the top five models. The models included are FAHR-FaceAge, MobileNet224 VMAGE, SENet50 VMAGE, MobileNet V3-large VMAGE and MobileNet96 VMAGE.

**d,** MAE for the Harvard RT patient dataset after the same fine-tuning, shown as a bar plot. The line plot displays the performance for each 5-year age bin for the top five models. The models included are FaceAge, FAHR-FaceAge, SENet50 IMDB-WIKI, DenseNet IMDB-WIKI and SENet50 VMAGE.

**e,** Analysis of plastic surgery procedures' effects on age estimation. Box plots compare pre- and post-surgery age estimates for five procedures: eyebrow correction (n=126 pairs, mean difference=0.29 years), eyelid correction (n=130 pairs, mean difference=-0.45 years), facelift (n=96 pairs, mean difference=-0.40 years), facial bones correction (n=105 pairs, mean difference=1.03 years), and nose correction (n=174 pairs, mean difference=1.11 years). Statistical analysis revealed significant changes only for nose correction (P<0.001), while other procedures showed non-significant differences (all P>0.05).

**f,** Assessment of makeup effects on age estimation (n=26 pairs). Box plots display age estimates before and after makeup application. Analysis showed non-significant changes in age estimation (mean difference=1.09 years, P>0.05).

**g,** Evaluation of viewing conditions effects. Box plots show age estimates across different presentation variables: viewing angles (22.5° left vs. frontal: mean difference=0.25 years; 22.5° right vs. frontal: mean difference=0.55 years; n=200 pairs), facial expressions (neutral vs. smiling: mean difference=0.79 years, n=200 pairs), and illumination conditions (light vs. dark: mean difference=0.55 years, n=181 pairs). All variations produced non-significant differences (all P>0.05).

**(e,f,g)**, The Wilcoxon signed-rank test was used to determine significance for each paired comparison, with the following annotation: ns (P>0.05), * (P≤0.05), ** (P≤0.01), *** (P≤0.001).

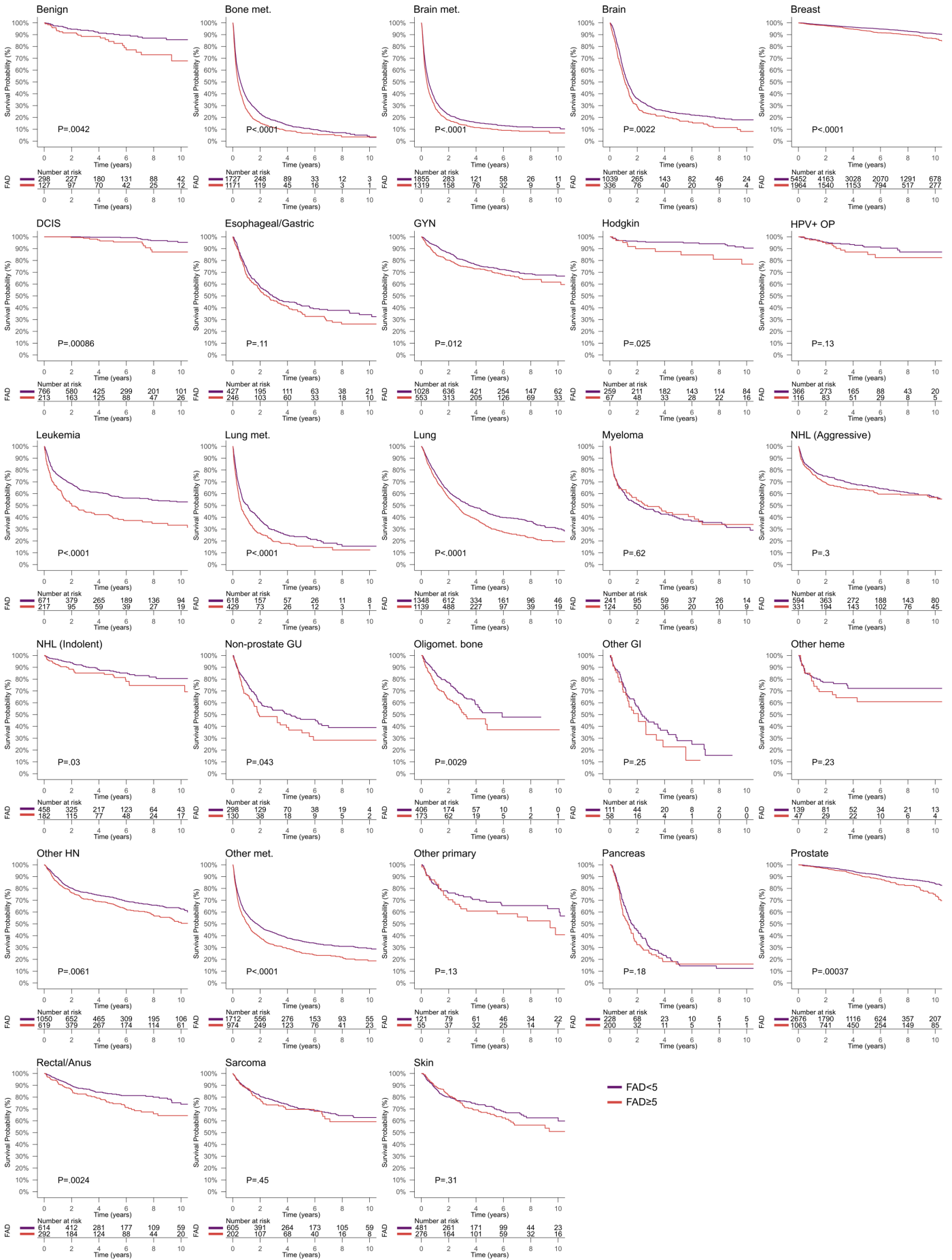

**Extended Data Figure 2: Kaplan-Meier Survival Curves by Cancer Type with FAD ≥ 5 and FAD < 5 on the Harvard RT Dataset**

Kaplan-Meier survival curves for each cancer type, stratified by FAHR-FaceAge deviation (FAD) with the split at FAD ≥ 5 and FAD < 5. The plots demonstrate the survival probabilities over time for patients within each cancer type based on their FAD value. For each cancer type, the survival curves illustrate that patients with an FAD ≥ 5 generally exhibit lower survival probabilities compared to those with an FAD < 5, highlighting the prognostic value of FAD as a biomarker across various cancer types. The significance of the difference between the groups is assessed using the log-rank test (P-values). met., metastases; DCIS, ductal carcinoma in situ; GYN, gynecological; HPV, human papillomavirus; OP, oropharyngeal; NHL, non-Hodgkin lymphoma; GU, genitourinary; GI, gastrointestinal; HN, head and neck.

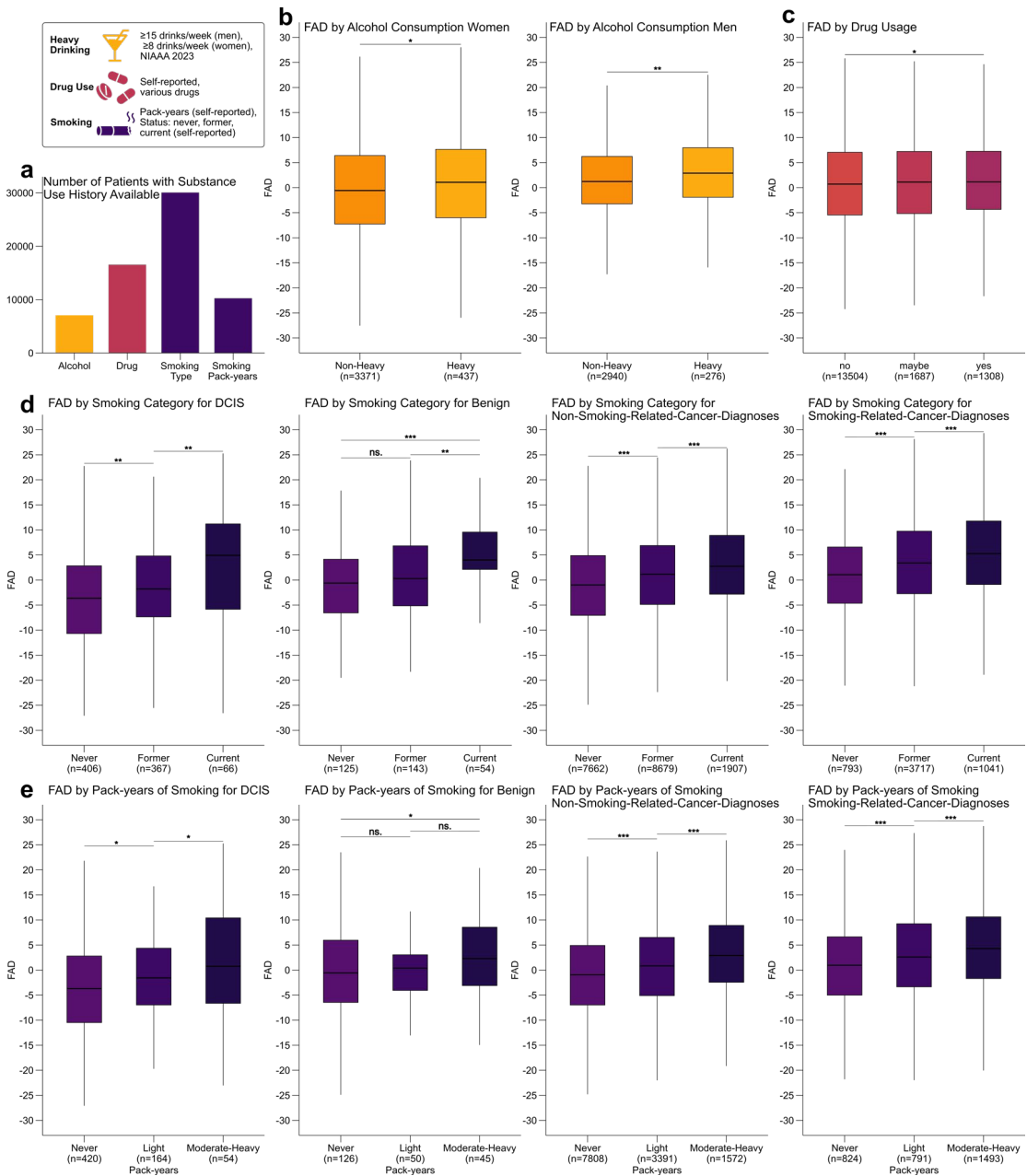

**Extended Data Figure 3: Impact of Lifestyle Factors on FAHR-FaceAge Deviation (FAD) on the Harvard RT Dataset**

**a,** Bar plot showing the number of patients with self-reported data on smoking, alcohol consumption, and drug use. This panel illustrates the amount of available data for each lifestyle factor.

**b,** Boxplots showing FAD grouped by heavy alcohol use, defined by NIAAA 2023 guidelines as consuming 15 or more drinks per week for men and 8 or more drinks per week for women. The plots show the distribution of FAD for heavy drinkers versus non-heavy drinkers for both men and women.

**c,** Boxplots showing FAD grouped by self-reported drug use categorized into three groups: No, Maybe, and Yes. This panel shows the distribution of FAD based on drug use.

**d,** Boxplots showing FAD grouped by self-reported smoking status: Never, Former, and Current.

**e,** Boxplots of FAD grouped by self-reported pack-years of cigarette consumption into never, light and heavy smokers. The plots show a statistically significant trend indicating that increased cigarette consumption is associated with a higher FAD. The definition of smoking-related and non-smoking-related cancers is as follows: Smoking-related cancers include Lung, Other HN, Non-prostate GU, and Pancreas. Non-smoking-related cancers include Breast, Prostate, Skin, HPV+ OP, NHL Indolent, HL (Hodgkin Lymphoma), Rectal/Anal, NHL Aggressive, Leukemia, Myeloma, Other Heme, Gyn, Sarcoma, Other GI, Other primary, Esophageal/Gastric, Brain Malignant, Brain, Hodgkin, Rectal/Anus, NHL (Aggressive), Other heme, NHL (Indolent), and GYN. Metastatic cancers were excluded. A relative FAHR-FaceSurvival of ≥3 was used to define smoking-related cancers[27]. The pack-year groups are defined as follows:

"Never" (0 pack-years), "Light" (0 < pack-years ≤ 20), "Moderate" (20 < pack-years ≤ 40), and "Heavy" (pack-years > 40)[28].

(**b,c,d,e**), The Wilcoxon rank-sum test was used to determine significance for each comparison, with the following annotation for significance: ns (P>0.05), * (P≤0.05), ** (P≤0.01), *** (P≤0.001).

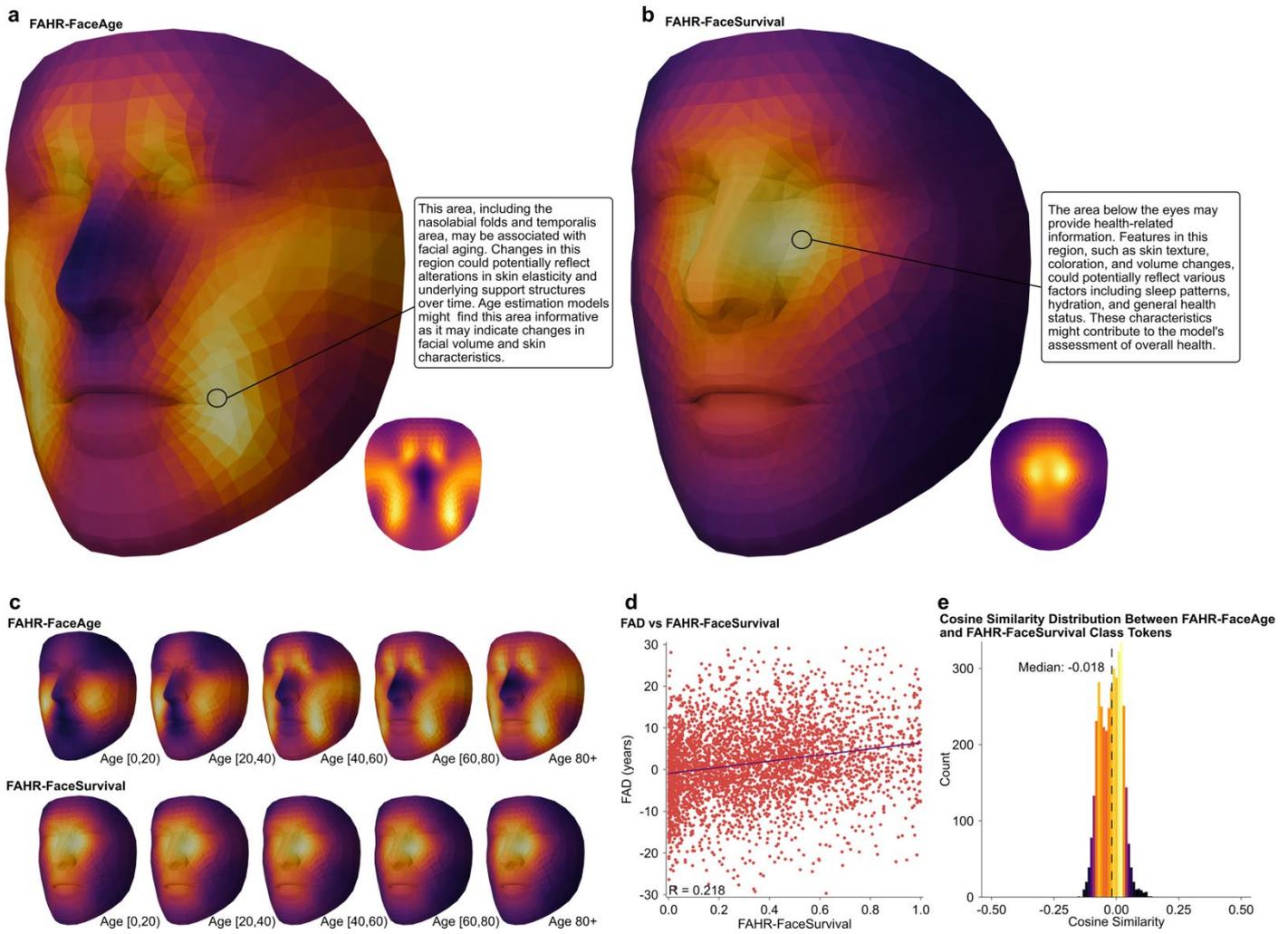

**Extended Data Figure 4: Attention Maps and Correlation Analysis for FAHR-FaceAge and FAHR-FaceSurvival Models on the Harvard RT Test Dataset**

**a,b,** Attention maps projected onto 3D and 2D canonical face models for FAHR-FaceAge (a) and FAHR-FaceSurvival (b). The heat maps highlight areas of focus for each model's predictive task. FAHR-FaceAge concentrates on regions associated with age-related facial changes particular on nasolabial folds and temporalis area. FAHR-FaceSurvival emphasizes features under the eyes.

**c,** Age-stratified attention maps for both FAHR-FaceAge and FAHR-FaceSurvival models.

**d,** Scatter plot showing the relationship between FAD and FAHR-FaceSurvival. The plot reveals a weak positive correlation (R = 0.218) between the two biomarkers. FAD, FAHR-FaceAge deviation (FAHR-FaceAge minus chronological age

**e,** Distribution of cosine similarities between FAHR-FaceAge and FAHR-FaceSurvival class token embeddings. The median similarity is -0.018.

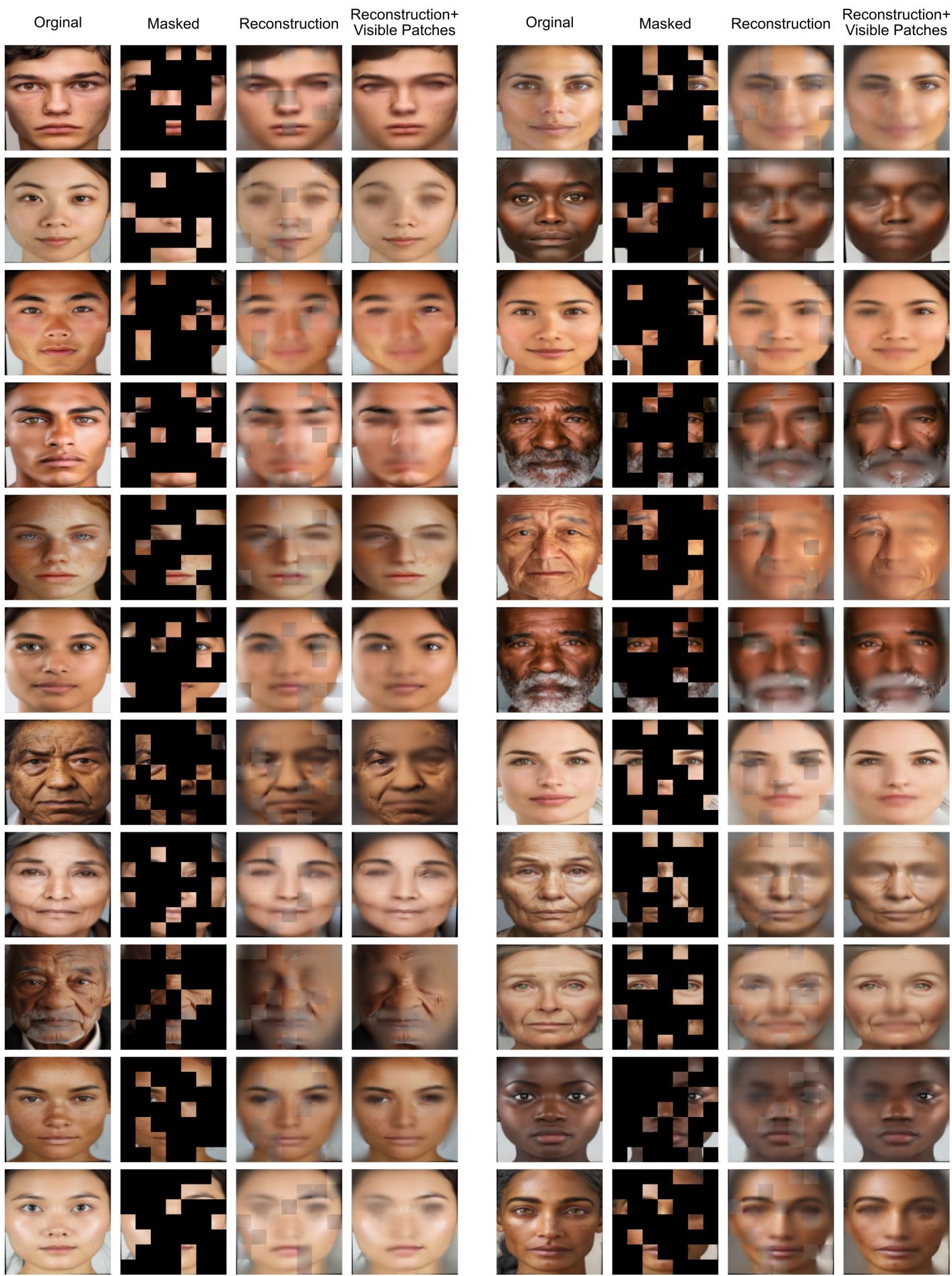

**Extended Data Figure 5: Reconstruction Ability of the Masked Autoencoder FAHR-Face Foundation**

This figure demonstrates the reconstruction ability of the masked autoencoder (MAE). Synthetic faces were generated using Midjourney with the following prompt: "Create a realistic face photo of a person, with the following attributes randomly selected: race, age, and sex. The individual can be of any ethnic background, any age from a child to an elderly person, and any sexes. The face should have natural expressions and be set in different lighting conditions and backgrounds. Ensure only one person is in the photo, and the appearance should be realistic and natural." All photos that contained exactly one face were used.

The images show the original synthetic faces and their reconstructions by the masked autoencoder across various ages, sex, and ethnic backgrounds. The reconstructions demonstrate the masked autoencoder's ability to accurately capture and recreate key facial features and structures, maintaining the integrity of the original faces. The performance of the masked autoencoder highlights its robustness and generalizability across diverse demographic characteristics.

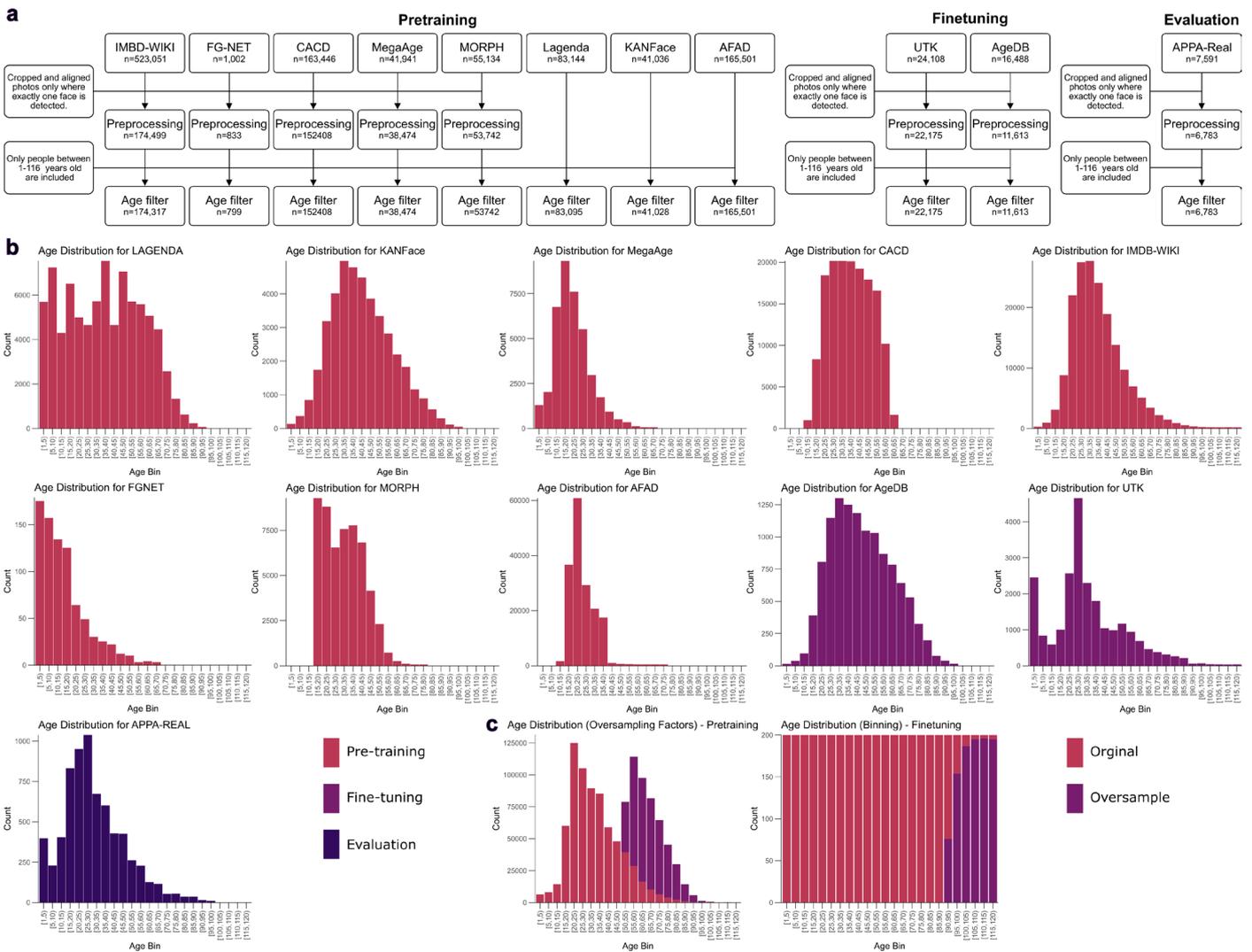

**Extended Data Figure 6: Age Dataset Collection, Distribution, and Oversampling**

**a,** Flowchart illustrating the collection and preprocessing of public datasets with age labels. The datasets include IMDB-WIKI, FG-NET, CACD, MegaAge, MORPH, Lagenda, KANFace, and AFAD for pre-training; UTK and AgeDB for fine-tuning; and APPA-Real for evaluation. The preprocessing steps involve cropping and aligning photos where exactly one face is detected, and filtering individuals aged between 1 and 116 years. The final number of photos used from each dataset is shown.

**b,** Age distribution of all used datasets. Each subplot shows the age distribution for a specific dataset used in pre-training (IMDB-WIKI, FG-NET, CACD, MegaAge, MORPH, Lagenda, KANFace, AFAD), fine-tuning (UTK, AgeDB), and evaluation (APPA-Real). The distributions illustrate the range of ages covered by each dataset, highlighting the diversity and representation across different age groups.

**c,** Illustration of the oversampling strategy applied to the datasets. The left panel shows the age distribution for the pre-training datasets before and after oversampling, while the right panel shows the same for the fine-tuning datasets. Oversampling is used to balance the age distribution, ensuring sufficient representation across all age groups and enhancing the model's ability to generalize.

**Extended Data Table 1. Patient demographics and characteristics of Harvard RT Patient Dataset and Maastro RT Dataset**

| Characteristics | Harvard RT Dataset | | | Maastro RT Dataset (n=4,892) |
|---|---|---|---|---|
| | FAHR-FaceAge (N=38,211) | FAHR-FaceSurvival (N=34,389) - training | FAHR-FaceSurvival (N=3,822) - test | |
| Age (years), median (range) | 64.1 (0.7-102.7) | 64.1 (0.7-102.7) | 64.1 (0.8 to 98.5) | 66.0 (22.0-94.0) |
| Number of events (deaths), No. (%) | 13769 (36.0) | 12357 (35.9) | 1412 (36.9) | 1811 (37.0) |
| Median follow-up time[a] (years) | 2.1 | 2.1 | 2.0 | 3.0 |
| FAHR-FaceAge (years), median (range) | 65.2 (-1.7 to 97.3) | 65.2 (-1.7 to 97.3) | 65.5 (-0.5 to 97.3) | 70.9 (12.0 to 97.0) |
| *FAD* (years) | | | | |
| Median (range) | 1.1 (-63.4 to 73.5) | 1.2 (-63.4 to 73.4) | 1.2 (-58.1 to 60.5) | 4.4 (-31.0 to 39.6) |
| ≥5, No. (%) | 12,623 (33.0) | 11,350 (33.0) | 1,273 (33.3) | 2,288 (46.8) |
| ≥10, No. (%) | 6,515 (17.1) | 5,837 (17.0) | 678 (17.7) | 1,103 (22.5) |
| ≤-5, No. (%) | 9,273 (24.3) | 8,418 (24.5) | 855 (22.4) | 410 (8.4%) |
| FAHR-FaceSurvival | | | | |
| Median (range) | | - | 0.3 (0.0-1.0) | 0.4 (0.0-1.0) |
| <0.25 | | - | 1,685 (44.1) | 1,405 (28.7) |
| 0.25-0.49 | | - | 1,126 (29.5) | 1,703 (34.8) |
| 0.5-0.74 | | - | 672 (17.6) | 1,461 (29.9) |
| ≥0.75 | | - | 339 (8.9) | 323 (6.6) |
| Sex, No. (%) | | | | |
| Female | 21,563 (56.4) | 19,385 (56.4) | 2,178 (57.0) | 2,309 (47.2) |
| Male | 16,648 (43.6) | 15,004 (43.6) | 1,644 (43.0) | 2,583 (52.8) |
| Race/ethnicity, No. (%) | | | | |
| White | 32,574 (87.5) | 29,317 (87.5) | 3,257 (87.6) | - |
| Black | 1,823 (4.9) | 1,641 (4.9) | 182 (4.9) | - |
| Asian | 1,239 (3.3) | 1,123 (3.4) | 116 (3.1) | - |
| Hispanic | 609 (1.6) | 550 (1.6) | 59 (1.6) | - |
| Other | 978 (2.7) | 872 (2.7) | 106 (2.8) | - |
| Unknown | 988 | 886 | 102 | - |
| Cancer risk group, No. (%) | | | | |
| Benign/DCIS | 1,404 (3.6) | 1,259 (3.7) | 145 (3.8) | - |
| Non-metastatic cancer | 26,423 (69.2) | 23,791 (69.2) | 2,632 (68.9) | - |
| Metastatic cancer | 10,384 (27.2) | 9.339 (27.2) | 1,045 (27.3) | - |
| Diagnosis, No. (%) | | | | |
| Benign | 425 (1.1) | 378 (1.1) | 47 (1.2) | - |
| DCIS | 979 (2.6) | 881 (2.6) | 98 (2.6) | - |
| Breast | 7,416 (19.4) | 6,709 (19.5) | 707 (18.5) | - |
| Prostate | 3,739 (9.8) | 3,356 (9.8) | 383 (10.0) | - |
| Non-prostate genitourinary | 428 (1.1) | 398 (1.2) | 30 (0.8) | - |
| Lung | 2,487 (6.5) | 2,245 (6.5) | 242 (6.3) | - |
| Esophageal/Gastric | 673 (1.8) | 613 (1.8) | 60 (1.6) | - |
| Pancreas | 428 (1.1) | 388 (1.1) | 40 (1.1) | - |
| Rectal/Anal | 906 (2.4) | 814 (2.4) | 92 (2.4) | - |
| Other gastrointestinal | 169 (0.4) | 150 (0.4) | 19 (0.5) | - |
| Gynecological | 1,581 (4.1) | 1,426 (4.2) | 155 (4.1) | - |
| HPV+ oropharynx | 482 (1.3) | 442 (1.3) | 40 (1.1) | - |
| Other head and neck | 1,669 (4.4) | 1,477 (4.3) | 192 (5.0) | - |
| Sarcoma | 807 (2.1) | 727 (2.1) | 80 (2.1) | - |
| Hodgkin lymphoma | 326 (0.9) | 300 (0.9) | 26 (0.7) | - |
| Non-Hodgkin lymphoma (indolent) | 640 (1.7) | 574 (1.7) | 66 (1.7) | - |
| Non-Hodgkin lymphoma (aggressive) | 925 (2.4) | 810 (2.4) | 115 (3.0) | - |
| Multiple myeloma | 365 (1.0) | 331 (1.0) | 34 (0.9) | - |
| Leukemia | 888 (2.3) | 796 (2.3) | 92 (2.4) | - |
| Other hematological | 186 (0.5) | 169 (0.5) | 17 (0.4) | - |
| Skin | 757 (2.0) | 677 (2.0) | 80 (2.1) | - |
| Brain (malignant) | 1,375 (3.6) | 1,234 (3.6) | 141 (3.7) | - |
| Other primary | 176 (0.5) | 155 (0.5) | 21 (0.6) | - |
| Oligometastasis bone | 579 (1.5) | 546 (1.6) | 33 (0.9) | - |
| Bone metastasis | 2,898 (7.6) | 2,619 (7.6) | 279 (7.3) | - |
| Lung metastasis | 1,047 (2.7) | 938 (2.7) | 109 (2.9) | - |
| Brain metastasis | 3,174 (8.3) | 2,832 (8.2) | 342 (9.0) | - |
| Other metastasis | 2,686 (7.0) | 2,404 (7.0) | 282 (7.4) | - |
| Cancer site, No. (%) | | | | |
| Benign | 425 (1.1) | 378 (1.1) | 47 (1.2) | - |
| Breast | 8,395 (22.0) | 7,590 (22.1) | 805 (21.1) | - |
| Prostate | 3,739 (9.8) | 3,356 (9.8) | 383 (10.0) | - |
| Lung | 2,487 (6.5) | 2,245 (6.5) | 242 (6.3) | - |
| Gastrointestinal | 2,176 (5.7) | 1,965 (5.7) | 211 (5.5) | - |
| Gynecological | 1,581 (4.1) | 1,426 (4.2) | 155 (4.1) | - |
| Head and neck | 2,151 (5.6) | 1,919 (5.6) | 232 (6.1) | - |
| Sarcoma | 807 (2.1) | 727 (2.1) | 80 (2.1) | - |
| Hematological | 3,330 (8.7) | 2,980 (8.7) | 350 (9.2) | - |
| Skin | 757 (2.0) | 677 (2.0) | 80 (2.1) | - |
| Brain (malignant) | 1,375 (3.6) | 1,234 (3.6) | 141 (3.7) | - |
| Other primary | 604 (1.6) | 553 (1.6) | 51 (1.3) | - |
| Bone metastasis | 3,477 (9.1) | 3,165 (9.2) | 312 (8.2) | - |
| Brain metastasis | 3,174 (8.3) | 2,832 (8.2) | 342 (9.0) | - |
| Cancer site (Maastro), No. (%) | | | | |
| Breast | - | - | - | 1337 (27.3) |
| Lung | - | - | - | 734 (15) |
| Gastrointestinal | - | - | - | 999 (20.4) |
| Genitourinary | - | - | - | 838 (17.1) |
| Head and Neck | - | - | - | 455 (9.3) |
| Other | - | - | - | 529 (10.8) |
| Treatment course intent, No. (%) | | | | |
| Curative | 26,371 (69.0) | 23,773 (69.1) | 2,598 (68.0) | - |
| Oligometastasis ablation | 3,262 (8.5) | 2,953 (8.6) | 309 (8.1) | - |
| Palliative | 8,578 (22.5) | 7,663 (22.3) | 915 (23.9) | - |
| Year treated, No. (%) | | | | |
| 2008-2015 | 13,447 (35.2) | 12,112 (35.2) | 1,335 (34.9) | - |
| 2016-2023 | 24,764 (64.8) | 22,277 (64.8) | 2,487 (65.1) | - |
| Radiation technique, No. (%) | | | | |
| 3D | 18,870 (52.8) | 16,995 (52.9) | 1,875 (52.4) | - |
| IMRT | 10,366 (29.0) | 9,304 (28.9) | 1,062 (29.7) | - |
| SBRT | 1,787 (5.0) | 1,640 (5.1) | 147 (4.1) | - |
| SRS | 1,981 (5.5) | 1,768 (5.5) | 213 (6.0) | - |
| Brachytherapy | 1,309 (3.7) | 1,189 (3.7) | 120 (3.4) | - |
| Electrons | 864 (2.4) | 767 (2.4) | 97 (2.7) | - |
| Adaptive | 550 (1.5) | 486 (1.5) | 64 (1.8) | - |
| Unknown | 2,484 | 2,240 | 244 | - |

Abbreviations: FAD, FAHR-FaceAge deviation (FAHR-FaceAge minus chronological age); No. (%), number of patients (percentage); DCIS, ductal carcinoma in situ; HPV, human papillomavirus; RT, radiotherapy; 3D, three-dimensional conformal radiotherapy; IMRT, intensity-modulated radiotherapy; SBRT, stereotactic body radiotherapy; SRS, stereotactic radiosurgery.

aMedian follow-up time was calculated using the reverse Kaplan-Meier estimator by reversing the censoring indicator[29].

**Extended Data Table 2. Results for FAHR-FaceAge variables in multivariate Cox regression for survival within each subgroup by sex, race/ethnicity, cancer risk, prognosis group, cancer site, and treatment course intent**

| Subgroups for individual multivariate model | Univariate result for FAD [decade] HR (95% CI) | P value | Multivariate result for FAD [decade] HR (95% CI) | P value | Univariate result for FAD ≥5 years HR (95% CI) | P value | Multivariate result for FAD ≥5 years HR (95% CI) | P value | Univariate result for FAD ≤-5 years HR (95% CI) | P value | Multivariate result for FAD ≤-5 years HR (95% CI) | P value |
|---|---|---|---|---|---|---|---|---|---|---|---|---|
| **Sex[a]** | | | | | | | | | | | | |
| Female | 1.36 (1.33-1.39) | <0.001 | 1.16 (1.13-1.18) | <0.001 | 1.77 (1.69-1.86) | <0.001 | 1.26 (1.19-1.32) | <0.001 | 0.61 (0.58-0.65) | <0.001 | 0.78 (0.73-0.83) | <0.001 |
| Male | 1.19 (1.56-1.23) | <0.001 | 1.11 (1.07-1.14) | <0.001 | 1.32 (1.26-1.39) | <0.001 | 1.17 (1.11-1.24) | <0.001 | 0.88 (0.83-0.94) | <0.001 | 0.89 (0.83-0.95) | <0.001 |
| **Age[b]** | | | | | | | | | | | | |
| <40 | 1.12 (1.03-1.22) | 0.006 | 1.12 (1.02-1.23) | 0.018 | 1.33 (1.11-1.58) | 0.002 | 1.44 (1.20-1.74) | <0.001 | 0.89 (0.77-1.04) | 0.146 | 0.97 (0.81-1.15) | 0.695 |
| 40-60 | 1.30 (1.26-1.34) | <0.001 | 1.17 (1.13-1.21) | <0.001 | 1.55 (1.44-1.66) | <0.001 | 1.26 (1.17-1.36) | <0.001 | 0.65 (0.61-0.71) | <0.001 | 0.79 (0.73-0.86) | <0.001 |
| 60-80 | 1.30 (1.26-1.33) | <0.001 | 1.15 (1.12-1.18) | <0.001 | 1.50 (1.43-1.57) | <0.001 | 1.25 (1.19-1.31) | <0.001 | 0.72 (0.68-0.77) | <0.001 | 0.84 (0.78-0.89) | <0.001 |
| ≥80 | 1.13 (1.06-1.19) | <0.001 | 1.14 (1.07-1.21) | <0.001 | 1.11 (1.01-1.22) | 0.025 | 1.13 (1.02-1.24) | 0.020 | 0.83 (0.73-0.94) | 0.002 | 0.78 (0.68-0.89) | <0.001 |
| **Race/ethnicity[c]** | | | | | | | | | | | | |
| White | 1.31 (1.28-1.33) | <0.001 | 1.15 (1.13-1.18) | <0.001 | 1.53 (1.47-1.58) | <0.001 | 1.23 (1.19-1.28) | <0.001 | 0.70 (0.66-0.73) | <0.001 | 0.83 (0.79-0.87) | <0.001 |
| Black | 1.15 (1.07-1.23) | <0.001 | 1.01 (0.94-1.08) | 0.863 | 1.37 (1.15-1.63) | <0.001 | 0.94 (0.78-1.14) | 0.539 | 0.84 (0.71-0.99) | 0.044 | 0.95 (0.80-1.13) | 0.566 |
| Asian | 1.27 (1.17-1.38) | <0.001 | 1.18 (1.07-1.30) | 0.001 | 1.69 (1.33-2.15) | <0.001 | 1.32 (1.01-1.71) | 0.041 | 0.51 (0.41-0.63) | <0.001 | 0.61 (0.48-0.77) | <0.001 |
| Hispanic* | 1.15 (1.01-1.32) | 0.041 | 1.13 (0.98-1.31) | 0.093 | 1.33 (1.00-1.75) | 0.047 | 1.29 (0.96-1.73) | 0.093 | 0.59 (0.43-0.81) | 0.001 | 0.68 (0.48-0.95) | 0.026 |
| Other | 1.37 (1.27-1.47) | <0.001 | 1.05 (0.94-1.16) | 0.415 | 1.64 (1.41-1.91) | <0.001 | 1.10 (0.86-1.41) | 0.462 | 0.58 (0.48-0.70) | <0.001 | 0.82 (0.63-1.07) | 0.150 |
| **Cancer risk groups[d]** | | | | | | | | | | | | |
| Benign/DCIS* | 1.69 (1.35-2.12) | <0.001 | 1.39 (1.09-1.78) | 0.009 | 2.91 (1.85-4.58) | <0.001 | 1.94 (1.21-3.12) | 0.006 | 0.42 (0.23-0.77) | 0.005 | 0.66 (0.36-1.21) | 0.178 |
| Non-metastatic cancer | 1.28 (1.25-1.32) | <0.001 | 1.18 (1.15-1.22) | <0.001 | 1.46 (1.39-1.54) | <0.001 | 1.29 (1.22-1.36) | <0.001 | 0.66 (0.62-0.70) | <0.001 | 0.75 (0.70-0.81) | <0.001 |
| Metastatic cancer | 1.16 (1.14-1.19) | <0.001 | 1.13 (1.11-1.16) | <0.001 | 1.31 (1.25-1.37) | <0.001 | 1.21 (1.15-1.27) | <0.001 | 0.83 (0.78-0.88) | <0.001 | 0.85 (0.80-0.91) | <0.001 |
| **Cancer site[d]** | | | | | | | | | | | | |
| Breast | 1.37 (1.25-1.50) | <0.001 | 1.28 (1.15-1.41) | <0.001 | 1.63 (1.34-1.98) | <0.001 | 1.36 (1.11-1.67) | 0.003 | 0.58 (0.46-0.72) | <0.001 | 0.63 (0.49-0.81) | <0.001 |
| Prostate | 1.37 (1.17-1.59) | <0.001 | 1.27 (1.08-1.50) | 0.004 | 1.54 (1.21-1.95) | <0.001 | 1.42 (1.09-1.86) | 0.009 | 0.72 (0.52-1.01) | 0.057 | 0.77 (0.54-1.11) | 0.165 |
| Lung | 1.18 (1.11-1.25) | <0.001 | 1.17 (1.09-1.25) | <0.001 | 1.30 (1.17-1.45) | <0.001 | 1.34 (1.19-1.51) | <0.001 | 0.68 (0.58-0.80) | <0.001 | 0.78 (0.66-0.93) | 0.005 |
| Gastrointestinal | 1.23 (1.14-1.32) | <0.001 | 1.14 (1.05-1.23) | 0.003 | 1.35 (1.18-1.55) | <0.001 | 1.10 (0.95-1.28) | 0.203 | 0.69 (0.58-0.82) | <0.001 | 0.75 (0.62-0.91) | <0.001 |
| Gynecological | 1.10 (1.00-1.22) | 0.052 | 1.05 (0.94-1.16) | 0.394 | 1.32 (1.06-1.63) | 0.012 | 1.17 (0.93-1.47) | 0.180 | 0.85 (0.67-1.09) | 0.195 | 0.88 (0.68-1.14) | 0.320 |
| Head and neck | 1.23 (1.12-1.36) | <0.001 | 1.13 (1.02-1.25) | 0.019 | 1.42 (1.20-1.68) | <0.001 | 1.30 (1.08-1.56) | 0.006 | 0.82 (0.65-1.02) | 0.079 | 0.85 (0.66-1.08) | 0.177 |
| Sarcoma | 1.03 (0.89-1.20) | 0.669 | 0.96 (0.83-1.11) | 0.537 | 1.12 (0.83-1.53) | 0.452 | 0.8 (0.62-1.21) | 0.399 | 1.10 (0.82-1.47) | 0.528 | 1.17 (0.85-1.61) | 0.333 |
| Hematological | 1.20 (1.12-1.28) | <0.001 | 1.12 (1.04-1.20) | 0.003 | 1.41 (1.24-1.59) | <0.001 | 1.23 (1.07-1.40) | 0.003 | 0.87 (0.75-1.02) | 0.080 | 0.88 (0.75-1.04) | 0.124 |
| Skin | 1.13 (0.96-1.34) | 0.156 | 1.42 (1.15-1.74) | 0.001 | 1.16 (0.87-1.53) | 0.314 | 1.61 (1.17-2.21) | 0.003 | 0.76 (0.51-1.13) | 0.170 | 0.62 (0.39-0.98) | 0.040 |
| Brain (malignant) | 1.18 (1.09-1.27) | <0.001 | 1.12 (1.04-1.22) | 0.005 | 1.25 (1.09-1.45) | 0.002 | 1.11 (0.96-1.30) | 0.168 | 0.76 (0.65-0.89) | <0.001 | 0.79 (0.67-0.93) | 0.004 |
| **Treatment course intent[e]** | | | | | | | | | | | | |
| Curative | 1.32 (1.28-1.35) | <0.001 | 1.16 (1.22-1.20) | <0.001 | 1.53 (1.45-1.62) | <0.001 | 1.24 (1.17-1.31) | <0.001 | 0.63 (0.59-0.67) | <0.001 | 0.79 (0.73-0.85) | <0.001 |
| Oligomet. ablation | 1.21 (1.15-1.27) | <0.001 | 1.15 (1.09-1.21) | <0.001 | 1.32 (1.20-1.46) | <0.001 | 1.19 (1.07-1.32) | 0.001 | 0.80 (0.71-0.90) | <0.001 | 0.84 (0.74-0.95) | 0.005 |
| Palliative | 1.13 (1.10-1.16) | <0.001 | 1.11 (1.08-1.14) | <0.001 | 1.22 (1.17-1.29) | <0.001 | 1.19 (1.13-1.25) | <0.001 | 0.82 (0.77-0.88) | <0.001 | 0.85 (0.80-0.91) | <0.001 |

Abbreviations: FAD, FAHR-FaceAge deviation (FAHR-FaceAge minus chronological Age); HR, hazard ratio; CI, confidence interval; DCIS, ductal carcinoma in situ; oligomet., oligometastasis.
[a]Multivariate analysis with age, race, cancer site, treatment course intent, year treated, radiation technique incorporated
[b]Multivariate analysis with sex, race, cancer site, treatment course intent, year treated, radiation technique incorporated
[c]Multivariate analysis with age, sex, cancer site, treatment course intent, year treated, radiation technique incorporated; *multivariate analysis with age, sex, cancer site, treatment course intent, year only for Hispanic to avoid overfitting in setting of limited number of events
[d]Multivariate analysis with age, sex, race, treatment course intent, year treated, radiation technique incorporated; *multivariate analysis with age, sex, treatment course intent only for benign/DCIS to avoid overfitting in setting of limited number of events
[e]Multivariate analysis with age, sex, race, cancer site, year treated, radiation technique incorporated

**Extended Data Table 3. Results for FAHR-FaceSurvival in univariate and multivariate Cox regression for overall survival within each subgroup by sex, race/ethnicity, cancer risk, prognosis group, cancer type, and treatment course intent for the Harvard RT dataset**

| Subgroups | Univariate for FAHR-FaceSurvival (0.1) | | Multivariate result for FAHR-FaceSurvival (0.1) | | Univariate result for FAHR-FaceSurvival (≥0.5vs<0.5) | | Multivariate result for FAHR-FaceSurvival (≥0.5vs<0.5) | |
|---|---|---|---|---|---|---|---|---|
| | HR (95% CI) | P value | HR (95% CI) | P value | HR (95% CI) | P value | HR (95% CI) | P value |
| Sex[a] | | | | | | | | |
| Female | 1.35 (1.32-1.39) | <0.001 | 1.17 (1.14-1.20) | <0.001 | 4.46 (3.86-5.16) | <0.001 | 2.14 (1.82-2.52) | <0.001 |
| Male | 1.28 (1.25-1.32) | <0.001 | 1.18 (1.14-1.22) | <0.001 | 2.89 (2.89-3.37) | <0.001 | 1.91 (1.62-2.26) | <0.001 |
| Age[b] | | | | | | | | |
| <40* | 1.18 (1.10-1.27) | <0.001 | 1.11 (1.02-1.20) | 0.018 | 1.98 (1.31-2.98) | 0.001 | 1.22 (0.73-2.04) | 0.450 |
| 40-60 | 1.35 (1.30-1.39) | <0.001 | 1.20 (1.15-1.25) | <0.001 | 5.12 (4.15-6.31) | <0.001 | 2.42 (1.92-3.06) | <0.001 |
| 60-80 | 1.34 (1.31-1.38) | <0.001 | 1.18 (1.14-1.21) | <0.001 | 3.51 (3.03-4.06) | <0.001 | 1.92 (1.64-2.25) | <0.001 |
| ≥80* | 1.21 (1.13-1.29) | <0.001 | 1.15 (1.07-1.23) | <0.001 | 2.28 (1.69-3.08) | <0.001 | 2.04 (1.49-2.79) | <0.001 |
| Race/ethnicity[c] | | | | | | | | |
| White | 1.33 (1.30-1.36) | <0.001 | 1.18 (1.15-1.21) | <0.001 | 3.82 (3.41-4.28) | <0.001 | 2.05 (1.81-2.32) | <0.001 |
| Black* | 1.41 (1.28-1.56) | <0.001 | 1.27 (1.14-1.42) | <0.001 | 5.38 (3.13-9.24) | <0.001 | 3.59 (1.98-6.51) | <0.001 |
| Asian* | 1.32 (1.17-1.49) | <0.001 | 1.29 (1.14-1.47) | <0.001 | 4.44 (2.17-9.09) | <0.001 | 3.79 (1.74-8.28) | <0.001 |
| Hispanic* | 1.14 (0.97-1.35) | 0.106 | 5.35 (0.93-30.73) | 0.060 | 1.91 (0.78-4.69) | 0.157 | 2.09 (0.84-5.23) | 0.115 |
| Other* | 1.29 (1.15-1.44) | <0.001 | 1.20 (1.06-1.35) | 0.003 | 1.92 (1.03-3.58) | 0.039 | 2.21 (1.23-3.99) | 0.008 |
| Cancer risk groups[d] | | | | | | | | |
| Benign/DCIS* | 1.17 (0.99-1.40) | 0.072 | -- | | 2.83 (0.73-10.97) | 0.132 | -- | -- |
| Non-metastatic cancer | 1.33 (1.29-1.37) | <0.001 | 1.26 (1.22-1.30) | <0.001 | 3.62 (3.10-4.23) | <0.001 | 2.85 (2.40-3.38) | <0.001 |
| Metastatic cancer | 1.19 (1.15-1.22) | <0.001 | 1.16 (1.12-1.19) | <0.001 | 2.19 (1.89-2.54) | <0.001 | 1.87 (1.60-2.19) | <0.001 |
| Cancer site[d] | | | | | | | | |
| Breast* | 1.31 (1.16-1.49) | <0.001 | 1.26 (1.10-1.43) | <0.001 | 3.86 (1.64-9.06) | 0.002 | 3.02 (1.27-7.16) | 0.012 |
| Prostate* | 1.46 (1.20-1.77) | <0.001 | 1.41 (1.15-1.72) | <0.001 | 4.03 (1.90-8.55) | <0.001 | 3.80 (1.68-8.62) | 0.001 |
| Lung* | 1.18 (1.09-1.29) | <0.001 | 1.18 (1.08-1.29) | <0.001 | 1.76 (1.23-2.51) | 0.002 | 1.75 (1.19-2.57) | 0.005 |
| Gastrointestinal* | 1.20 (1.11-1.31) | <0.001 | 1.10 (1.00-1.21) | 0.055 | 2.41 (1.54-3.77) | <0.001 | 1.86 (1.15-3.01) | 0.011 |
| Gynecological* | 1.21 (1.06-1.39) | 0.006 | 1.08 (0.93-1.25) | 0.330 | 1.86 (0.85-4.08) | 0.121 | 0.83 (0.36-1.92) | 0.657 |
| Head and neck* | 1.27 (1.15-1.41) | <0.001 | 1.21 (1.08-1.35) | <0.001 | 3.39 (2.00-5.75) | <0.001 | 2.57 (1.45-4.57) | 0.001 |
| Sarcoma* | 1.20 (1.01-1.43) | 0.039 | 1.10 (0.92-1.32) | 0.293 | 3.48 (1.61-7.56) | 0.002 | 3.48 (1.55-7.81) | 0.003 |
| Hematological* | 1.29 (1.20-1.38) | <0.001 | 1.30 (1.21-1.39) | <0.001 | 2.31 (1.63-3.29) | <0.001 | 2.63 (1.82-3.80) | <0.001 |
| Skin* | 1.30 (1.05-1.61) | 0.016 | -- | -- | 2.73 (1.03-7.24) | 0.044 | -- | -- |
| Brain (malignant)* | 1.09 (1.02-1.17) | 0.013 | 1.09 (1.01-1.18) | 0.033 | 1.44 (0.96-2.16) | 0.082 | 1.27 (0.80-2.02) | 0.312 |
| Treatment course intent[e] | | | | | | | | |
| Curative | 1.27 (1.23-1.31) | <0.001 | 1.13 (1.08-1.17) | <0.001 | 2.95 (2.47-3.52) | <0.001 | 1.67 (1.37-2.04) | <0.001 |
| Oligomet. Ablation* | 1.23 (1.16-1.31) | <0.001 | 1.18 (1.10-1.25) | <0.001 | 3.06 (2.23-4.20) | <0.001 | 2.31 (1.64-3.25) | <0.001 |
| Palliative | 1.18 (1.15-1.21) | <0.001 | 1.18 (1.14-1.21) | <0.001 | 2.08 (1.79-2.41) | <0.001 | 2.00 (1.70-2.34) | <0.001 |

Abbreviations: FAHR-FaceSurvival, fair artificial intelligence for health recognition-survival prediction score based on facial photographs; cont, continuous; cat, categorized; HR, hazard ratio; CI, confidence interval; DCIS, ductal carcinoma in situ; oligomet., oligometastasis

[a]Multivariate analysis with age, race, cancer site, treatment course intent, year treated, radiation technique incorporated

[b]Multivariate analysis with sex, race, cancer site, treatment course intent, year treated, radiation technique incorporated; *multivariate analysis with age, sex, race, treatment course intent only for age<40 and age≥80 to avoid overfitting in setting of limited number of events

[c]Multivariate analysis with age, sex, cancer site, treatment course intent, year treated, radiation technique incorporated; *multivariate analysis with age, sex, treatment course intent only for Black/others; age and sex only for Asian/Hispanic, to avoid overfitting in setting of limited number of events

[d]Multivariate analysis with age, sex, race, treatment course intent, year treated, radiation technique incorporated; *no multivariate model generated for benign/DCIS and skin due to limited number of events; multivariate analysis with age, sex, race, treatment course intent only for favorable, lung, gastrointestinal, hematological, and brain; age and treatment course intent only for breast, prostate, gynecological, and sarcoma; age, sex, treatment course intent only for head and neck, to avoid overfitting in setting of limited number of events

[e]Multivariate analysis with age, sex, race, cancer site, year treated, radiation technique incorporated; multivariate analysis with age, sex, and cancer site only for oligomet. ablation, to avoid overfitting in setting of limited number of events

**Extended Data Table 4. Diagnosis grouping according to ICD-9 or ICD-10 codes**

| ICD diagnosis | Diagnosis | Cancer site | ICD-9 codes | ICD-10 codes |
|---|---|---|---|---|
| Benign brain | Benign | Benign | 225.1, 225.2, 225.9, 225, 192.1, 747.81 | D32.9, D36.10, D36.11, D32.0, D33.3, Q28.2, D42.0, D35.2, D44.4 |
| Benign cardiovascular | Benign | Benign | 447.1 | I25.110, I73.9, I70.1, I70.212, I70.292, I70.90, I70.8, I47.2, I77.0, I74.4 |
| Benign HO/OA | Benign | Benign | 728.13, 213.6, 213.9, 215.9, 213.7, 215.3, 715.15, 728.1 | M89.8X9, M61.00, M61.49, M61.9, M61.50, M61.551, D16.11, M61.452, M16.12, M61.49, S73.004S, M25.8, M89.8X0, M61.552 |
| Benign keloid | Benign | Benign | 701.4 | L73.0, L91.0 |
| Benign other | Benign | Benign | 705.83 | D21.12, N62, M72.0, D21.21, D11.0, L73.2 |
| Bone met. | Bone met. | Bone met. | 198.5 | C79.51, C79.52 |
| Brain met. | Brain met. | Brain met. | 198.3 | C79.31, C79.32 |
| Other met. | Other met. | Other met. | 198.8, 196.9, 196.3, 198.2, 198.6, 198.7, 198.89, 196.2, 198.81, 198.1, 198.4, 196.6, 196.5, 198.82, 196.8 | C77.4, C77.3, C77.9, C77.5, C79.89, C79.40, C79.82, C78.4, C77.2, G95.20, C79.9, C79.49, C77.0, C77, C79.81, C79.2, C78.89, C79.72, C79.70, C78.6, C79.11, C76.3 |
| DCIS/LCIS[a] | DCIS | Breast | 233 | D05.12, D05.11, D05.01, D05.02, D05.81, D05.10, D05.80, D05.82, D05.90, D05.91, D05.92 |
| GI other prim.[a] | Other GI | GI | 153.3, 153.9, 153.6 | C18.4, C18.7, C76.2, C18.9, C78.5, C18.2, C18.1, C18.8 |
| GU other prim.[a] | Non-prostate GU | Other prim. | 186.9, 187.4, 187.2, 187.3 | C68.0, C62.12, C62.11, C60.9, C60.2 |
| GYN other prim.[a] | GYN | GYN | NA | C56.9, C56.1, C57.4, C57.00, C56.2 |
| Heme other[a] | Other Heme | Heme | 284.9, 238.75, 238.73, 284.8, 289.83, 238.79, 238.76, 284.09 | D61.9, D46.9, D75.81, D47.1, C94.6, D61.89, C96.A, D61.3, C88.0 |
| Hodgkin lymphoma[a] | Hodgkin lymphoma | Heme | 201.02, 201.41, 201.42, 201.44, 201.48, 201.4, 201.51, 201.52, 201.53, 201.55, 201.58, 201.5, 201.62, 201.64, 201.68, 201.6, 201.91, 201.92, 201.98, 201.9 | C81.01, C81.02, C81.04, C81.05, C81.08, C81.10, C81.11, C81.12, C81.13, C81.14, C81.18, C83.11, C83.12, C83.13, C81.14, C83.18, C81.20, C81.21, C81.24, C81.40, C81.70, C81.71, C81.72, C81.78, C81.79, C81.90, C81.91, C81.92, C81.94, C81.95, C81.96, C81.98, C81.99, C81.00 |
| HPV- OP SCC[a/b] | Other HN | HN | 146.1, 146.2, 146.4, 146.5, 146.7, 146.6, 146.8, 146.9, 146 | C10.2, C10.3, C10.9, C10.8, C09.0, C09.1, C09.9, C09.8, C01 |
| Leukemia | Leukemia | Heme | 183, 204, 204.01, 204.02, 204.1, 204.21, 204.8, 205, 205.01, 205.02, 205.1, 205.11, 205.12, 205.21, 205.9, 206.01, 208.01, 208.02, 207.82 | C91.00, C91.01, C91.02, C92.00, C92.01, C92.02, C92.50, C92.51, C92.62, C92.A0, C92.A1, C92.A2, C93.00, C95.00, C95.01, C95.02, C91.51, C92.0, C91.0, C95.92, C92.90, C95.91 |
| Liver met. | Other met. | Other met. | 197.7 | C78.7 |
| Lung/thoracic met. | Lung met. | Other met. | 197.1, 197.2, 196.1, 197 | C78.00, C78.01, C78.02, C77.1, C78.2, C78.1, C78.39 |
| Male prim. breast[a] | Other prim. | Other prim. | 175, 175.90 | C50.929, C50.121, C50.921, C50.122, C50.922 |
| Melanoma[a] | Skin | Skin | 172.9, 172.3, 172.4, 172.5, 172.6, 172.7, 172.8, 172.2 | C43.11, C43.21, C43.30, C43.31, C43.39, C43.51, C43.59, C43.60, C43.61, C43.62, C43.72, C43.8, C43.9, C43.4 |
| Merkel cell[a] | Skin | Skin | 209.31, 209.32, 209.33, 209.34, 209.36 | C4A.10, C4A.11, C4A.112, C4A.22, C4A.30, C4A.31, C4A.39, C4A.4, C4A.51, C4A.52, C4A.59, C4A.61, C4A.60, C4A.62, C4A.70, C4A.71, C4A.72, C4A.8, C4A.9 |
| Merkel cell met. | Other met. | Other met. | 209.75 | C7B.1 |
| Multiple myeloma[a] | Multiple myeloma | Heme | 203, 203.01, 203.02, 238.6 | C90.00, C90.01, C90.02, C90.0, C90.30, C90.20 |
| NHL (aggressive) | NHL (aggressive) | Heme | 200.41, 200.48, 200.71, 200.78, 200.72, 200.75, 200.73, 200.28, 202.7, 200.7, 200.4, 200.28, 200.68 | C83.10, C83.11, C83.16, C83.18, C83.19, C83.30, C83.31, C83.32, C83.33, C83.34, C83.35, C83.38, C83.39, C84.45, C84.48, C84.49, C84.61, C84.65, C84.69, C84.79, C84.09, C84.A9, C84.A, C93.10, C84.A5 |
| NHL (unspecified)[c] | NHL (aggressive or indolent) | Heme | 202.8, 202.88, 202.81, 202.85, 202.83 | C83.52, C83.70, C85.19, C85.80, C85.81, C85.83, C85.84, C85.85, C85.88, C85.89, C85.90, C85.9, C85.91, C85.92, C85.93, C85.94, C85.95, C85.96, C85.98, C85.99 |
| NHL (indolent) | NHL (indolent) | Heme | 200.31, 200.33, 200.34, 202.05 | C82, C82.00, C82.01, C82.02, C82.03, C82.04, C82.05, C82.06, C82.07, C82.08, C82.09, C82.10, C82.11, C82.12, C82.13, C82.14, C82.15, C82.16, C82.17, C82.18, C82.19, C82.20, C82.21, C82.22, C82.23, C82.24, C82.25, C82.26, C82.27, C82.28, C82.29, C82.30, C82.31, C82.32, C82.33, C82.34, C82.35, C82.36, C82.37, C82.38, C82.39, C82.40, C82.41, C82.42, C82.43, C82.44, C82.45, C82.46, C82.47, C82.48, C82.49, C82.50, C82.51, C82.52, C82.53, C82.54, C82.55, C82.56, C82.57, C82.58, C82.59, C82.60, C82.61, C82.62, C82.63, C82.64, C82.65, C82.66, C82.67, C82.68, C82.69, C82.70, C82.71, C82.72, C82.73, C82.74, C82.75, C82.76, C82.77, C82.78, C82.79, C82.80, C82.81, C82.82, C82.83, C82.84, C82.85, C82.86, C82.87, C82.88, C82.89, C82.90, C82.91, C82.92, C82.93, C82.98, C82.99, C83.00, C83.01, C83.09, C83.14, C83.80, C83.81, C83.82, C83.89, C83.99, C84.00, C84.01, C84.10, C84.A0, C85.10, C88.4, C91.10, C91.12 |
| Other HN[a] | Other HN | HN | 141, 141.1, 141.2, 141.3, 141.4, 141.5, 141.6, 141.8, 141.9, 142, 142.1, 142.8, 142.9, 143.1, 144, 144.1, 144.8, 144.9, 145, 145.2, 145.3, 145.5, 145.6, 145.8, 145.9, 147.1, 147.2, 147.8, 147.9, 148.1, 148.3, 148.8., 148.9, 1489, 149.8, 160, 160.1, 160.2, 160.3, 160.4, 160.8, 160.9, 161, 161.1, 161.2, 161.3, 161.8, 161.9, 193, 149.9, 148.8, 210.2, 199.1, 140.1 | C00.0, C00.1, C00.4, C00.9, C02.0, C02.1, C02.2, C02.3, C02.4, C02.8, C02.9, C03.0, C03.1, C03.9, C04.0, C04.1, C04.8, C04.9, C05.0, C05.1, C05.2, C05.9, C06.0, C06.1, C06.2, C06.89, C06.9, C07, C08.0, C08.1, C08.9, C11.0, C11.1, C11.2, C11.8, C11.9, C12, C13.0, C13.2, C13.8, C13.9, C14.0, C14.8, C30.0, C31.0, C31.1, C31.8, C31.9, C32.0, C32.1, C32.8, C32.9, C73, C76.0, C85.819, C18.2, C32 |
| Other unspecified prim.[a] | Other prim. | Other prim. | 192.2, 194, 195, 170 | C7A.8, C7B.03, C47.9, C72.0, C26.1, C72.9, D48.9, C80.1, C70.1 |
| Prim. bladder/kidney[a] | Non-prostate GU | Other prim. | 188.1, 188.2, 188.3, 188.4, 188.5, 188.6, 188.8, 188.9, 189.1, 189.2, 189.3, 188, 189 | C67.0, C67.1, C67.2, C67.3, C67.4, C67.5, C64.1, C64.2, C64.9, C67.8, C67.9, C67.6, C74.02, C74.91, C68.9, C79.01, C79.71, C74.90, C65.1, C74.92, C74.01, C66.2, C74.00 |
| Prim. breast[a] | Breast | Breast | 174.1, 174.2, 174.3, 174.4, 174.5, 174.6, 174.8, 174.9, 174 | C50.912, C50.911, C50.411, C50.412, C50.812, C50.811, C50.111, C50.112, C50.212, C50.919, C50.511, C50.011, C50.211, C50.311, C50.611, C50.012, C50.312, C50.512, C50.612, C50.512, C50.612, C50.819 |
| Prim. esophageal/gastric[a] | Esophageal/Gastric | GI | 151.9, 151.3, 151.4, 151.5, 151.8, 151.1, 151.6, 150.9, 150.8, 150.5, 150.4, 150.3, 150.2, 150.1, 152, 150, 151 | C15.9, C15.5, C15.3, C15.4, C15.8, C16.2, C16.0, C16.1, C16.3, C16.4, C16.8, C16.9, C17.0 |
| Prim. GYN (cervix/vulva)[a] | GYN | GYN | 180.1, 180.2, 180.9, 180.8, 184.4, 184, 180 | C53.0, C53.1, C53.8, C53.9, C51.8, C51.9, C51.0, C52 |
| Prim. GYN uterine* | GYN | GYN | 182.8, 179, 182 | C54.1, C54.8, C54.9, C54.2, C54.3, C55 |
| Prim. liver/GB[a] | Other GI | GI | 156.9, 156.8, 156.1, 155.1, 155, 156 | C22.0, C22.1, C22.2, C22.8, C22.9, C24.0, C23, C24.8, C24.9, C22 |
| Prim. lung/thorax[a] | Lung | Lung | 162.2, 162.3, 162.4, 162.5, 162.8, 162.9, 163.8, 163.9, 164.9, 162, 164, 163, 165.8, 164.8 | C34.11, C34.12, C34.31, C34.32, C34.90, C34.91, C34.92, C34.01, C34.02, C34.10, C34.80, C34.81, C34.82, C38.1, C38.3, C38.4, C45.0, C76.1, C34.2, C34.30, C33, C34.00, C7A.090, C37, C34.1, C34.9, C45.7, R91.1 |
| Prim. pancreas[a] | Pancreas | GI | 157.1, 157.2, 157.8, 157.9, 156.2, 157 | C25.1, C25.2, C25.3, C25.4, C25.7, C25.8, C25.9, C24.1, C25.0, C25 |
| Prim. prostate[a] | Prostate | Prostate | 185 | C61 |
| Prim. rectum/anus[a] | Rectal/Anal | GI | 154.1, 154.2, 154.3, 154.8, 154 | C20, C19, C21.0, C21.1, C21.8, C21 |
| Primary HGG/GBM | Brain (malignant) | Brain (malignant) | 191.1, 191.2, 191.3, 191.4, 191.5, 191.6, 191.7, 191.8, 191.9, 191 | C71.0, C71.1, C71.2, C71.3, C71.4, C71.5, C71.6, C71.7, C71.8, C71.9, C70.0, C70.9 |
| Sarcoma[a] | Sarcoma | Sarcoma | 170.1, 170.2, 170.4, 170.6, 170.7, 170.9, 171, 171.2, 171.3, 171.4, 171.5, 171.6, 171.7, 171.8, 171.9 | C40.01, C40.20, C40.22, C40.21, C40.30, C40.90, C41.0, C41.1, C41.2, C41.3, C41.4, C41.9, C49.1, C49.10, C49.12, C49.2, C49.20, C49.21, C49.22, C49.3, C49.4, C49.5, C49.6, C49.8, C49.80, C49.81, C49.82, C76.52, C49.0, C48.0, C49.9, C40.32, C76.51, C49.11, C76.41, C48.2, C46.0 |
| SCC/BCC[a] | Skin | Skin | 173.32, 173.31, 173.3, 173.01, 173.02, 173.09, 173.1, 173.12, 173.2, 173.21, 173.22, 173.4, 173.41, 173.42, 173.49, 173.5, 173.52, 173.6, 173.61, 173.62, 173.7, 173.71, 173.72, 173.8, 173.82, 173.89, 173.9, 173.92, 173.99, 173.39 | C44.00, C44.01, C44.02, C44.111, C44.112, C44.121, C44.202, C44.209, C44.212, C44.219, C44.221, C44.222, C44.229, C44.292, C44.299, C44.300, C44.301, C44.309, C44.310, C44.311, C44.319, C44.320, C44.321, C44.329, C44.40, C44.41, C44.42, C44.509, C44.519, C44.520, C44.529, C44.590, C44.599, C44.602, C44.612, C44.619, C44.621, C44.622, C44.629, C44.719, C44.721, C44.722, C44.729, C44.80, C44.81, C44.89, C44.90, C44.91, C44.92, C44.99, C44.390, C44.391, C44.399 |

Abbreviations: ICD, international classification of diseases; HO/OA, heterotopic ossification/osteoarthritis; met., metastasis; DCIS, ductal carcinoma in situ; LCIS, lobular carcinoma in situ; GI, gastrointestinal; prim., primary (non-metastatic); GU, genitourinary; GYN, gynecological; NA, not applicable; Heme, hematological; HPV, human papillomavirus; OP, oropharynx; SCC, squamous cell carcinoma; NHL, non-Hodgkin lymphoma; HN, head and neck; GB, gall bladder; HGG, high grade glioma; GBM, glioblastoma; BCC, basal cell carcinoma

[a]Classified as metastasis if diagnosis description specifies "metastatic," "metastasis," "metastases," "mets," "met," or "stage IV"
[b]Classified as HPV+ OP if diagnosis description specifies "p16+," "p16 positive," "HPV positive," "HPV+," "+HPV," "HPV associated," "HPV related," or "HPV mediated"
[c]Classified as NHL (aggressive) if diagnosis description specifies "DLBCL," "large cell," "diffuse large," "anaplastic," "T cell," "mantle," "high grade," "G3," or "grade 3," versus NHL (indolent) if diagnosis description specified "follicular," "marginal zone," "indolent," "low grade," "G1," or "grade 1"